\documentclass[aps,prd,twocolumn,showpacs,amsmath,amssymb,nofootinbib,eqsecnum, preprintnumbers]{revtex4-1}
\usepackage{graphicx}
\usepackage{epstopdf}
\usepackage{epsfig}
\usepackage{xcolor}

\def\appendix{{\newpage\section*{Appendix}}\let\appendix\section%
        {\setcounter{section}{0}
        \gdef\thesection{\Alph{section}}}\section}

\newcommand{\be}{\begin{equation}}
\newcommand{\ee}{\end{equation}}
\newcommand{\bear}{\begin{eqnarray}}
\newcommand{\eear}{\end{eqnarray}}
\newcommand{\ba}{\begin{array}}
\newcommand{\ea}{\end{array}}

\begin{document}

\title{Configurational entropy and instability of tachyonic braneworld}
\author{Chong Oh Lee}
\email{cohlee@gmail.com}
\affiliation{Department of Physics, Kunsan
National University, Kunsan 573-701, Republic of Korea}

\begin{abstract}
We consider tachyonic braneworld with a bulk cosmological constant
and investigate a configurational entropy
of various magnitudes of scale factor. It is found that for a bulk negative/zero cosmological constant,
the configurational entropy has a global minimum
when the magnitude of scale factor reaches the critical value.
This result seems to have intriguing implications such that an accelerated rate of the universe
and cosmological inflation rate for radiation/matter domination are able to be determined by such critical value.
We also find that the configurational entropy almost monotonically decreases for a bulk positive cosmological constant
as the magnitude of scale factor grows up. We find an exact solution of tachyonic braneworld in a bulk de Sitter space.
It is shown that such system under scalar perturbations is stable for some constraint relation.
Furthermore, we also find that tachyonic braneworld model with a bulk negative/zero cosmological constant is always stable
under scalar perturbations.
\end{abstract}
\maketitle
\section{Introduction}
The D-brane with tachyon condensation~\cite{Sen:1998sm} has been
suggested in search for off-shell structure of string theory.
Tachyon dynamics on unstable D-branes  was  investigated
in~\cite{Sen:2002nu,Sen:2002in,Sen:2002an} through boundary
conformal field theory and effective field theory. In particular,
such effective field theory has been extensively applied to
inhomogeneous rolling tachyon~\cite{Cline:2002it,Felder:2002sv,
Felder:2004xu, Barnaby:2004nk},
creation of closed string and particle~\cite{Lambert:2003zr,Kluson:2003qk}
and instability of D-branes of codimension-one
~\cite{Lambert:2003zr,Kim:2003ina,Kim:2003ma, Sen:2003bc}.

A branch of models with bulk scalar fields has been recently suggested
in search for tachyonic braneworld  scenario with a bulk cosmological constant.
It has been found that there is an exact solution
for tachyonic braneworld model with/without a bulk cosmological
constant~\cite{German:2012rv, Barbosa-Cendejas:2017vgm} and
such model is stable under scalar perturbations~\cite{Barbosa-Cendejas:2017vgm, German:2015cna}.

The configurational entropy has been suggested in search for the informational
entropy~\cite{Gleiser:2011di} and
recently applied to the spatially localized energy solutions of nonlinear models
~\cite{Gleiser:2012tu}.
The configurational entropy has used to study
instability of a variety of physical systems~\cite{Gleiser:2012tu,Gleiser:2013mga,Gleiser:2014ipa,
Gleiser:2015rwa,Bernardini:2016hvx,Bernardini:2016qit,Casadio:2016aum,Braga:2016wzx,Lee:2017ero,Gleiser:2018kbq,Lee:2018zmp}.

The paper is organized as follows: in the following section, we will
calculate the configurational entropy of tachyon effective theory
in case of the Anti-de Sitter (AdS), flat and de Sitter (dS).
In the next section, we will explore instability of tachyonic braneworld
in the bulk dS space.
In the last section we will give our conclusion.

\section{Configurational entropy}
One considers the energy density $\rho(x)$ as the function of the position $x$
in $d$-dimensional space and the system is spatially localized energy.
Its Fourier transforms $\rho(k)$ is written as
\bear\label{den}{\textstyle
\rho(k)=\left(\frac{1}{\sqrt{2\pi}}\right)^d \int \rho(x)e^{-ik\cdot x}d^dx},
\eear
and the modal fraction reads
\bear\label{mf}{\textstyle
{\cal F}(k)=\frac{|{\cal P}(k)|^2}{\int|{\cal P}(k)|^2d^dk},}
\eear
which measures the relative weight of a given mode $k$.
One may define the configurational entropy $S_{\rm c}[{\cal F}]$ as
\bear{\textstyle
S_{\rm c}[{\cal F}]=-\sum_{l=1}^n {\cal F}_l \ln({\cal F}_l),}
\eear
which reduces to
\bear\label{CS}\textstyle
S_{\rm c}[{\cal F}]=-\int_{-\infty}^{\infty} {\cal G}(k) \ln [{\cal G}(k)]d^dk,
\eear
for $n\rightarrow\infty$,
where ${\cal G}(k)={\cal F}(k)/{\cal F}(k)_{\rm max}$ and the maximum modal fraction
${\cal F}(k)_{\rm max}$.

The action for the tachyonic braneworld model is given as
\bear\label{ac}{\textstyle
A=A_{c}-A_m,}
\eear
where
$A_{c}$ is the five-dimensional gravity action with the bulk cosmological constant $\Lambda_5$,
$A_m$ is the action of the matter in the bulk,
\bear\textstyle
A_{c}=
\int d^5x\sqrt{-\det (g_{\mu\nu})}\left(\frac{1}{2k_5^2}{\cal R}-\Lambda_5\right),
\eear
\bear{\textstyle
A_{m}=\int d^5x V(T) \sqrt{-\det (g_{\mu\nu}+\partial_{\mu}T\partial_{\nu}T)}.}
\eear
Here $k_5$ is the five-dimensional gravitational constant ($k_5^2\equiv8\pi G_5$),
${\cal R}$ is the five-dimensional scalar curvature, $T$ is the tachyon field,
and $V(T)$ is the  tachyon potential. The action (\ref{ac}) leads to
the five-dimensional Einstein equation with the bulk cosmological constant
\bear\label{ee}{\textstyle
G_{\mu\nu}=-k_5^2\Lambda_5 g_{\mu\nu}+k_5^2T_{\mu\nu}.}
\eear
The five-dimensional metric ansatz with an induced 3-brane
of Friedmann-Robertson-Walker(FRW) background type is given as
\bear\label {FRW}\scriptstyle
ds^2 = e^{2f(x)}[-dt^2+a^2(t)(d\alpha^2+d\beta^2+d\gamma^2)+dx^2],
\eear
where $e^{2f(x)}$ denotes the warp factor, and $a(t)$ does the scale factor on the brane.
Then, the Einstein tensor $G_{\mu\nu}$ (\ref{ee}) have the nonvanishing components:
\bear\textstyle
G_{tt}=\frac{3 \dot{a}^2}{a^2}-3\left(f''+2f'^2\right),
\eear
\bear\textstyle
G_{\alpha\alpha}&=&G_{\beta\beta}=G_{\gamma\gamma}\nonumber\\
&=&-2\ddot{a}a-\dot{a}^2+3a^2\left(f''+2f'^2\right),
\eear
\bear\textstyle
G_{xx}=-3\left(\frac{\ddot{a}}{a}+\frac{\dot{a}^2}{a^2}\right)+6f'^2,
\eear
where the dot and the prime denote time and spatial derivative, respectively.
After taking $a(t)=e^{Ht}$, the matter field equations are given as
\bear\label{fe1}\textstyle
&&T''-f'T'+4f'T'(e^{-2f}T'^2+1)\nonumber\\
&&=(T'^2+e^{2f})\frac{\partial_T V(T)}{V(T)},
\eear
\bear\label{fe2}\textstyle
f''-f'^2+H^2=-k_5^2\frac{V(T)T'^2}{3\sqrt{e^{-2f}T'^2+1}},
\eear
\bear\label{fe3}\textstyle
f'^2-H^2+\frac{k_5^2 e^{2f} \Lambda_5}{6}=-k_5^2\frac{e^{2f}V(T)}{6\sqrt{e^{-2f}T'^2+1}},
\eear
where $H$ is constant.

After evaluating Eqs. (\ref{fe1}), (\ref{fe2}), and (\ref{fe3}),
the potential $V(T)$ is explicitly obtained as \cite{German:2012rv,Barbosa-Cendejas:2017vgm,German:2015cna}
\begin{align}\label{VT}
{\scriptscriptstyle V(T)=\left\{
\begin{array}{cl}
&{\scriptscriptstyle -\Lambda_5 {\rm sech}\left[\sqrt{-\frac{2}{3}k_5^2\Lambda_5}T\right]
\sqrt{6{\rm sech}^2\left[\sqrt{-\frac{2}{3}k_5^2\Lambda_5}T\right]-1}}
~~~~~{\scriptscriptstyle(\Lambda_5<0)}\\
&\\
&{\scriptscriptstyle \frac{3\sqrt{6}\,\sigma^2}{k_5^2}\sec\big[2\,{\rm am}(i\sqrt{2}\,\sigma T,2)\big]}
~~~~~{\scriptscriptstyle(\Lambda_5=0,~n=\frac{1}{2})},\\
&\\
&{\scriptscriptstyle -\Lambda_5 {\rm sec}\left[\sqrt{\frac{2}{3}k_5^2\Lambda_5}T\right]
\sqrt{6{\rm sec}^2\left[\sqrt{\frac{2}{3}k_5^2\Lambda_5}T\right]-1}}
~~~~~{\scriptscriptstyle(\Lambda_5>0)}
\end{array}
\right.}
\end{align}
where $\sigma$ is constant, and Jacobi amplitude am$(u,l)$ is defined by
\bear\textstyle
\phi={\rm am}(u,l)=\int_{0}^{u}{\rm dn}(u',l')du'.
\eear
Here, dn$(u,l)$ is a Jacobi elliptic function with elliptic modulus.
After taking $k_5=1$ and $\sigma=1$, with various values of the cosmological constant $\Lambda_5$,
the potential $V(T)$ as the function of the tachyon field $T$ is depicted in Fig. 1.

\begin{figure}[!htbp]
\begin{center}
{\includegraphics[width=8cm]{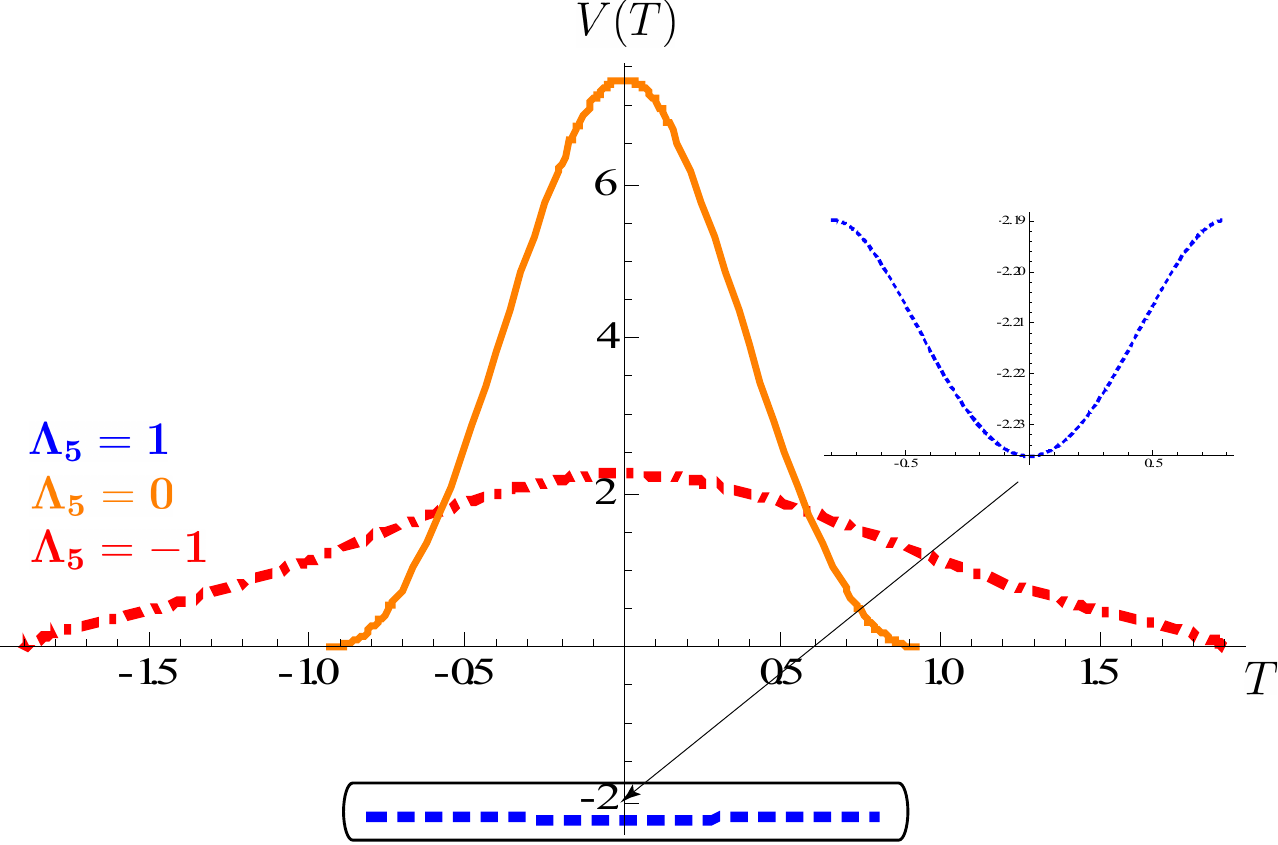}}
\end{center}
\vspace{-0.6cm}
\caption{{\footnotesize Plot of potential $V(T)$ as
the function of the tachyon field $T$
for $k_5=1$ and $\sigma=1$
(red dotted-dashed curve for $\Lambda_5 =-1$, orange solid curve for $\Lambda_5 =0$,
blue dashed curve for $\Lambda_5 =1$, respectively).}}
\label{figI-1}
\end{figure}

\subsection{$\Lambda_5<0$ case}
When we consider the five-dimensional gravity action with the negative cosmological constant $\Lambda_5<0$,
after evaluating Eqs. (\ref{fe1}), (\ref{fe2}), and (\ref{fe3}),
the warped factor $f(x)$ and the tachyon scalar field $T(x)$ are obtained as
\bear\textstyle
f(x)=\frac{1}{2}\ln[-\frac{6H^2{\rm sech}[H(2x+\lambda)]}{k_5^2 \Lambda_5}],
\eear
\bear\label{Tx}\textstyle
T(x)=\pm\sqrt{\frac{-3}{2k_5^2 \Lambda_5}} \tanh^{-1}
\left[\frac{\sinh\left[\frac{H(2x+\lambda)}{2}\right]}{\sqrt{\cosh[H(2x+\lambda)]}}\right],
\eear
and substituting into the potential $V(T)$ (\ref{VT}), the potential $V(x)$ is written as
\bear\scriptstyle
V(x)=-\Lambda_5 \frac{\sqrt{3{\rm sech}[H(2x+\lambda)]+2}\sqrt{{\rm sech}[H(2x+\lambda)]+1}}{\sqrt{2}},
\eear
where $\lambda$ is constant.
The energy density $\rho(x)$ is given as
\bear\textstyle
\rho(x)=\frac{12H^2(\cosh[H(2x+\lambda)]+2)}{k_5^2((\cosh[2H(2x+\lambda)]+1)}.
\eear

When $\Lambda_5=-1$, shapes of potential $V(x)$ as the function of the position $x$
for various values of $H$ are depicted in Fig. 2.
The bigger $H$ becomes, the more convex function in potential $V(x)$.
Then, profiles of tachyon filed  $T(x)$ for various $H$ and
profiles of energy density are depicted in Fig. 3 and Fig. 4, respectively.
The five-dimensional curvature scalar ${\cal R}$ is written as
\bear\textstyle
{\cal R}=-\frac{14}{3}k_5^2 \Lambda_5 {\rm sech}[H(2x+\lambda)]
\eear
which is positive definite and asymptotically flat along the coordinate $x$
as shown in Fig. 5.
The configurational entropy of tachyon filed $S_{\rm c, AdS}$ is numerically calculated
by Eqs. (\ref{den}), (\ref{mf}), and (\ref{CS}), and
is depicted in Fig. 6.
Especially, $H$ increases, the configurational entropy
reaches the minimum value ($S_{\rm c,AdS}=0.0160325$) at a critical point ($H_{c}=1.61$).

In fact, the singular modes in momentum space broadly spread out and
become plane waves with equally distributed modal. In addition
through Fourier transforms, the corresponding modes in position
space are sharply localized. The localized modes in position space
have the maximum configurational entropy due to large amount of
momentum modes whereas widespread modes in position space have the
minimum configurational entropy due to small amount. Furthermore,
configurational entropy is the portion of the entropy of a system
that is given as the discrete representative positions of its
constituent particles. For example, it may refer to the number of
spin configurations in a magnet. As a given energy increases, its
number increases. Thus, the smaller configurational entropy of
physical system becomes, the smaller its amount of energy to
generate its configurations. The larger configurational entropy of
physical system becomes, the larger its amount of energy to generate
its configurations. Thus, one may expect that the predominant
tachyonic states occurs at the minimum configurational entropy.

When the universe is accelerating expansion due to a negative
pressure fluid, the simple example is that the universe with the
scale factor $a(t)=e^{Ht}$ is driven exponentially toward a flat
geometry. Then the so-called horizon problem is solved by looking at
the conformal time since accelerating expansion implies that the
horizon size is shrinking in comoving units. Conformal time
(comoving horizon) is defined as \bear
\tau\equiv\int\frac{dt}{a(t)}=-\frac{1}{a(t)H}. \eear In particular,
it seems that the constant $H$ may be determined by the critical
point $H_c$. The five-dimensional metric ansatz (\ref{FRW}) on the
brane reduces to the flat FRW metric via conformal time $\tau$
\bear\label{FRW}\textstyle
ds_4^2=a^2(\tau)(-d\tau^2+d\alpha^2+d\beta^2+d\gamma^2), \eear where
the constant warp factor $e^{2f_0}$ is able to be absorbed by a
rescaling of the spatial coordinates. For a
radiation-dominated/matter-dominated universe the evolution of the
scale factor $a(\tau)$ in the metric (\ref{FRW}) is obtained solving
the Friedmann equations:
\begin{align}
a(\tau)\propto\left\{
\begin{array}{cl}
&\tau=-\frac{1}{a(t)H}~~~~~{(\rm RD)},\\
&\\
&\tau^2=\frac{1}{a^2(t)H^2}~~~~~{(\rm MD)}
\end{array}
\right.
\end{align}
which seems to be determined by the above critical point $H_c$. Thus
since the predominant tachyonic states occurs at $H_c$, it seems
that accelerated rate of the universe and cosmological inflation
rate for radiation/matter domination are able to be determined by
such critical value.
\begin{figure}[!htbp]
\begin{center}
{\includegraphics[width=8cm]{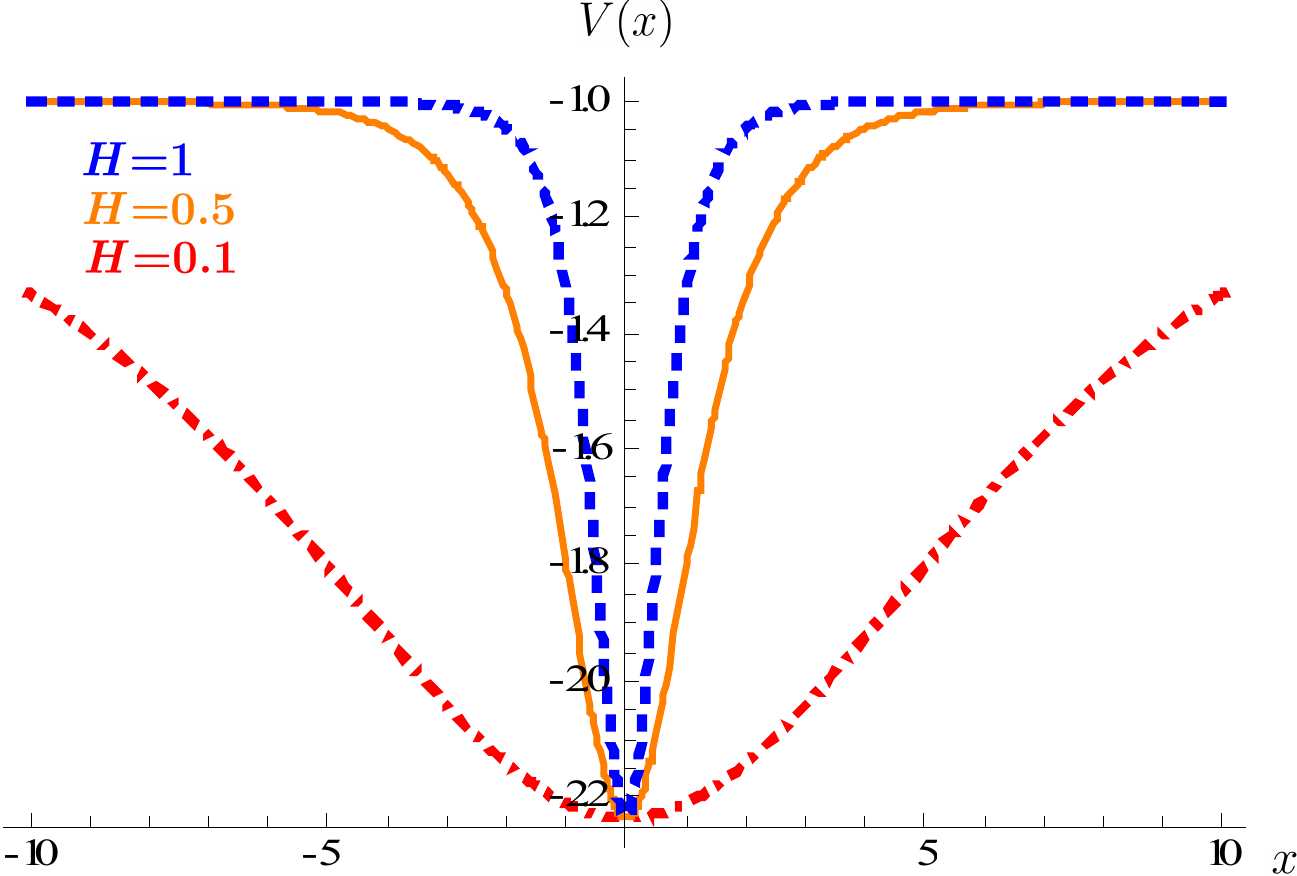}}
\end{center}
\vspace{-0.6cm}
\caption{{\footnotesize Plot of potential $V(x)$ as
the function of the position $x$
for $k_5=1$, $\Lambda_5=-1$, and $\lambda=0$
(red dotted-dashed curve for $H =0.1$, orange solid curve for $H=0.5$,
blue dashed curve for $H=1$,
respectively).}}
\label{figI}
\end{figure}

\begin{figure}[!htbp]
\begin{center}
{\includegraphics[width=8cm]{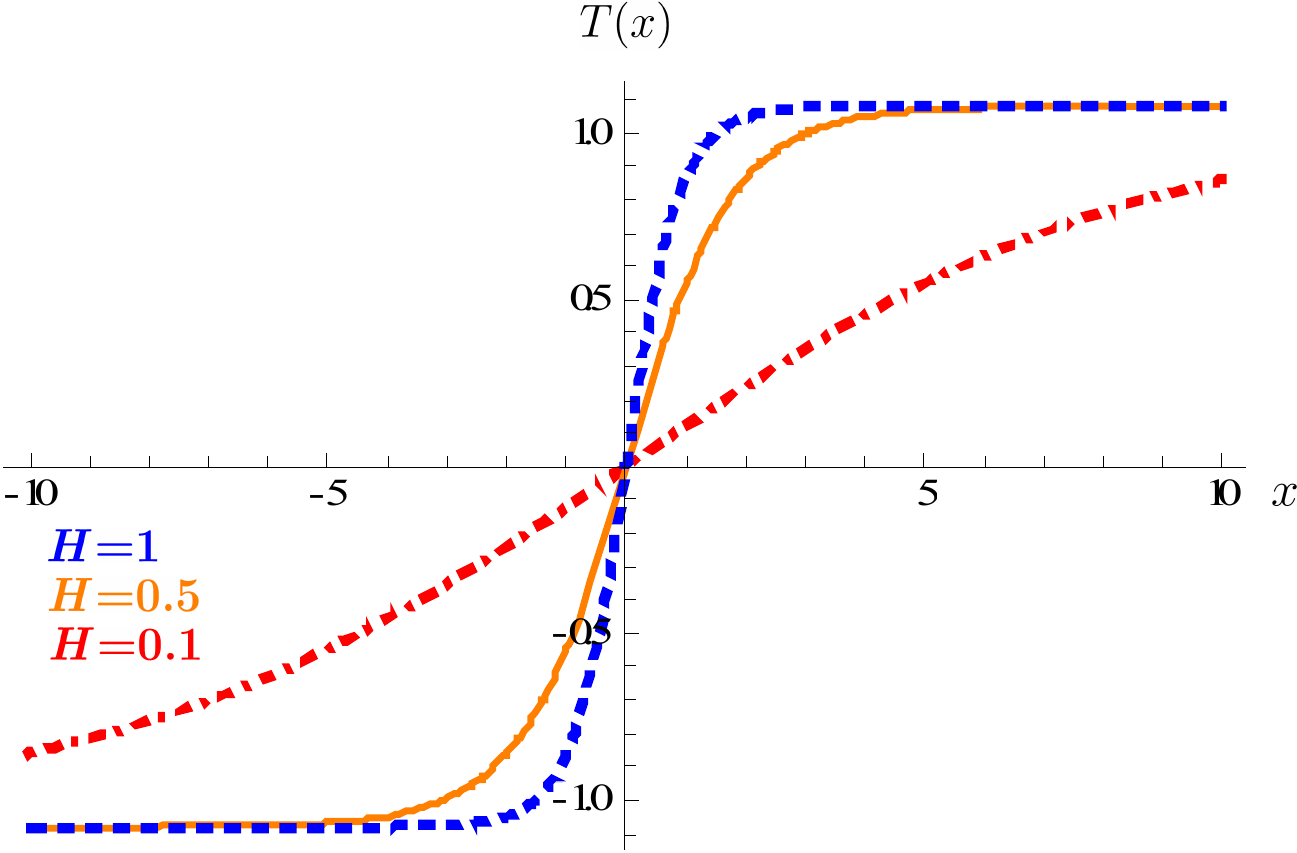}}
\end{center}
\vspace{-0.6cm}
\caption{{\footnotesize Plot of the tachyon field $T(x)$ as
the function of the position $x$
for $k_5=1$, $\Lambda_5=-1$, and $\lambda=0$
(red dotted-dashed curve for $H =0.1$, orange solid curve for $H=0.5$,
blue dashed curve for $H=1$,
respectively).}}
\label{figII}
\end{figure}

\begin{figure}[!htbp]
\begin{center}
{\includegraphics[width=8cm]{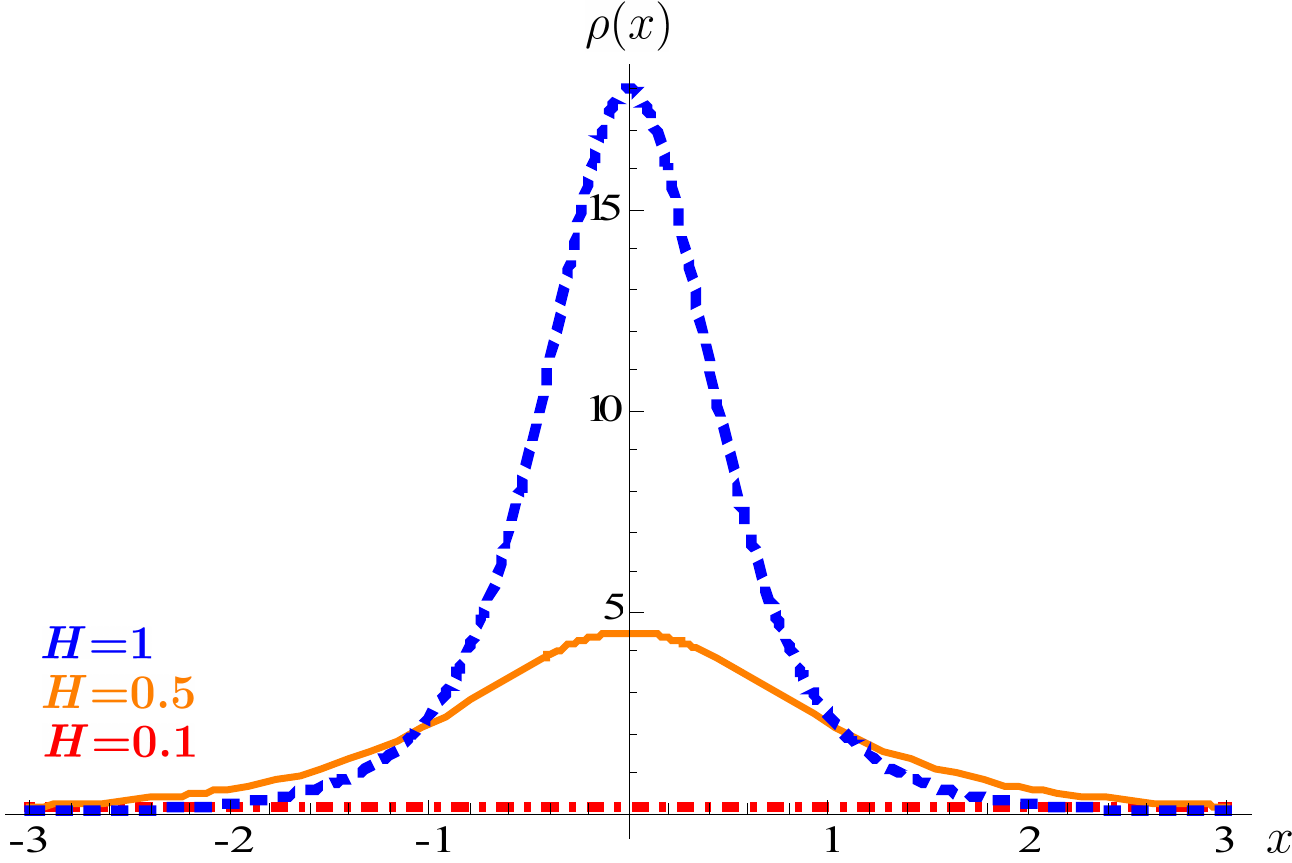}}
\end{center}
\vspace{-0.6cm}
\caption{{\footnotesize Plot of energy density $\rho(x)$ as
the function of the position $x$
for $k_5=1$, $\Lambda_5=-1$, and $\lambda=0$
(red dotted-dashed curve for $H =0.1$, orange solid curve for $H=0.5$,
blue dashed curve for $H=1$,
respectively).}}
\label{figIII}
\end{figure}

\begin{figure}[!htbp]
\begin{center}
{\includegraphics[width=8cm]{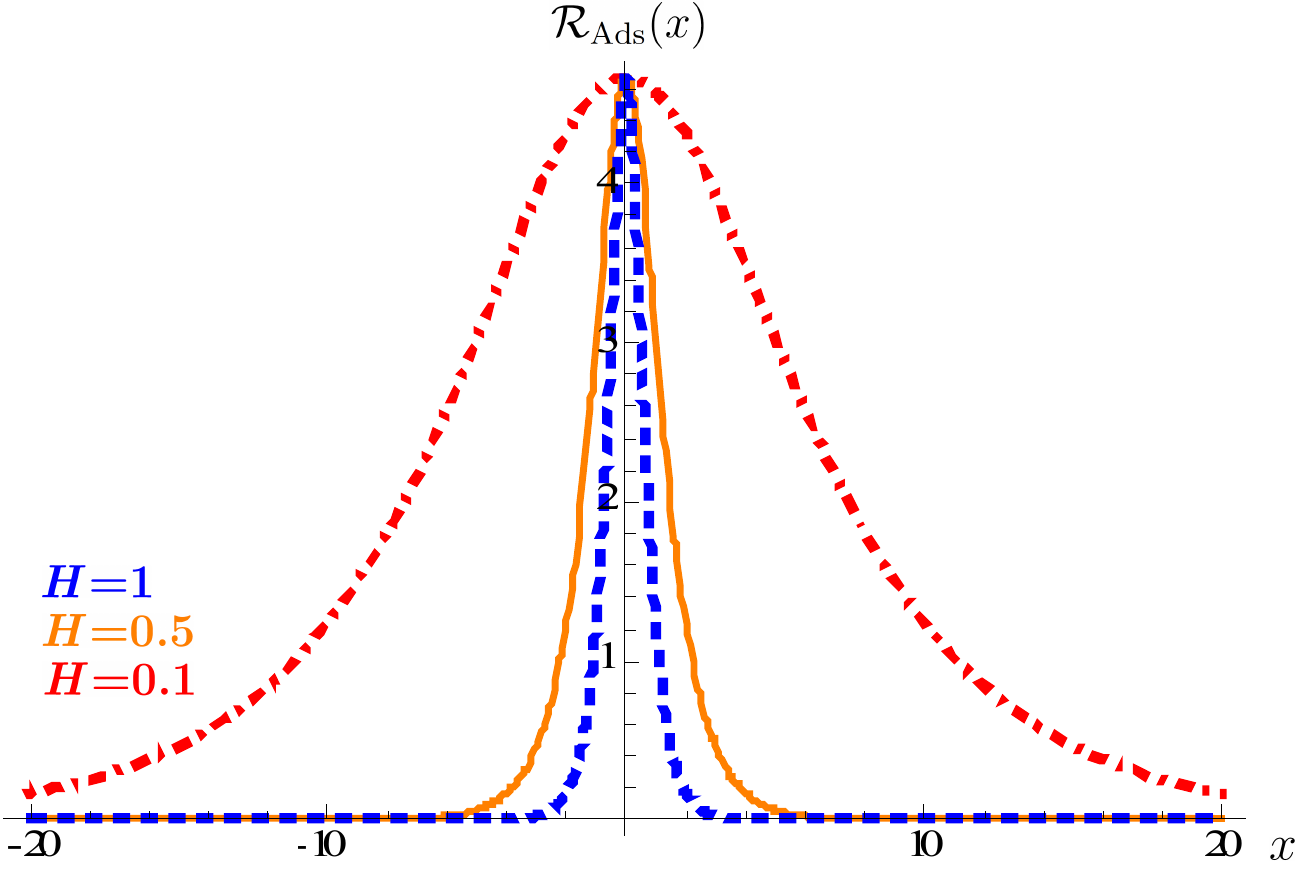}}
\end{center}
\vspace{-0.6cm}
\caption{{\footnotesize Plot of curvature sclar ${\cal R}_{\rm AdS}(x)$ as
the function of the position $x$
for $k_5 =1$, $\Lambda_5=-1$, and $\lambda=0$
(red dotted-dashed curve for $H =0.1$, orange solid curve for $H=0.5$,
blue dashed curve for $H=1$,
respectively).}}
\label{fig1}
\end{figure}

\begin{figure}[!htbp]
\begin{center}
{\includegraphics[width=8cm]{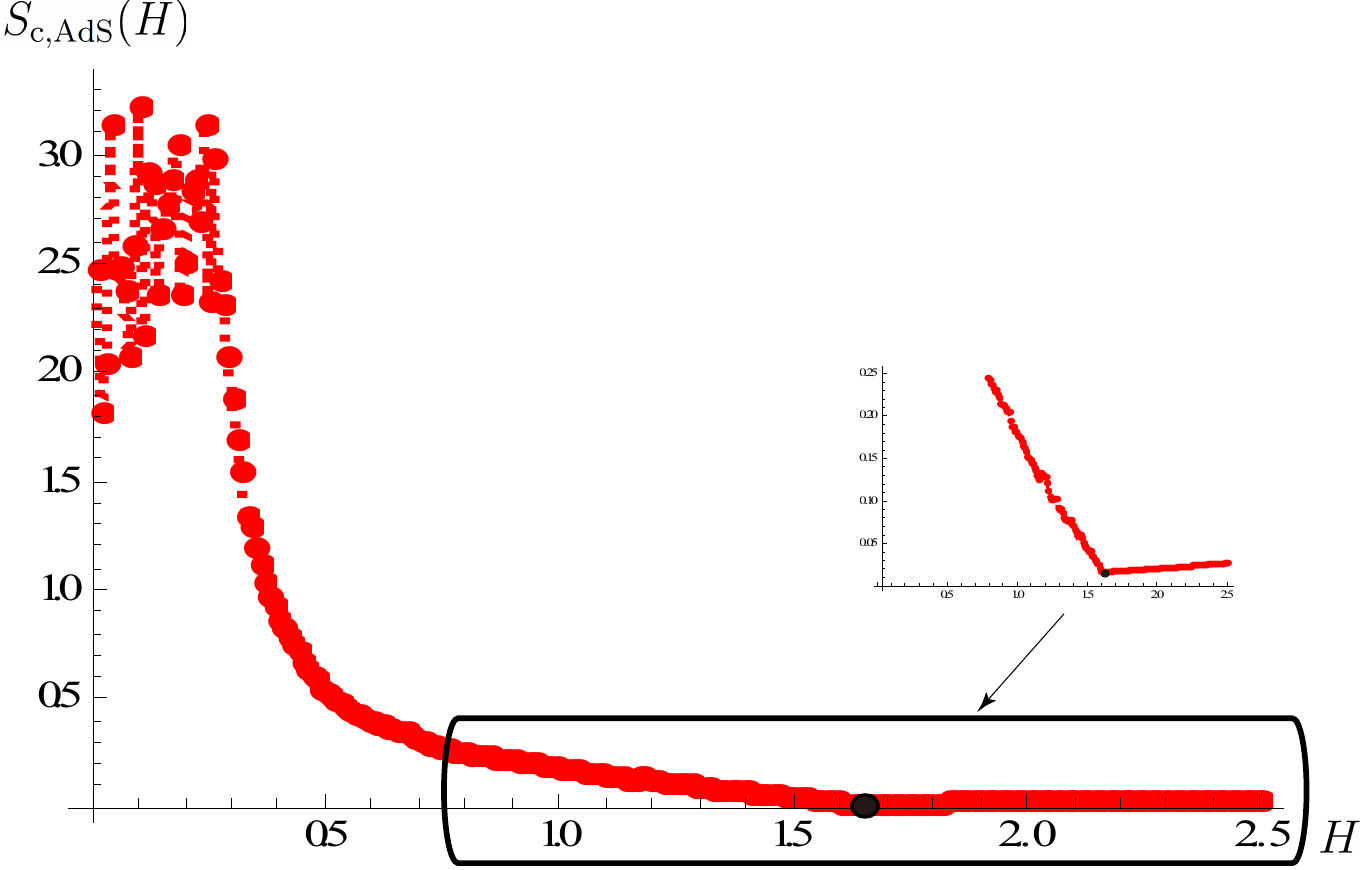}}
\end{center}
\vspace{-0.6cm}
\caption{{\footnotesize Plot of tachyon configurational entropy $S_{\rm c,AdS}$
as the function of $H$ for $k_5=1$, $\Lambda_5=-1$, and $\lambda=1$.}}
\label{figIV}
\end{figure}

\subsection{$\Lambda_5=0$ case}
As discussed in the previous section we will apply a
similar analysis to the five-dimensional gravity action without a
bulk cosmological constant $\Lambda_5$.

The potential $V(x)$ is obtained as \cite{Barbosa-Cendejas:2017vgm}
\bear\label{V0}\textstyle
V(x)=\frac{3\sqrt{\frac{2(n+1)}{n}}\,\sigma^{4n}}{k_5^2 H^{4n-2}}{\rm sech}^{2(1-n)}\left[H\left(\frac{x}{n}+\lambda\right)\right],
\eear
where $\sigma$ is constant.
The tachyon field $T(x)$ is
\bear\label{T0}{\scriptstyle
T(x)=\pm\frac{\sqrt{\frac{n}{2(n-1)}}{}_2F_1(\frac{1}{2},\frac{1-n}{2};\frac{3-n}{2};\cosh[H(\frac{x}{n}+\lambda)])}
{\sigma^{2n}H^{1-2n}\cosh^{n-1}[H(\frac{x}{n}+\lambda)]}+k},
\eear
where the hypergeometric function ${}_2F_1(a,b;c;z)$ is defined for $|z| > 1$ by the power series
\bear{\scriptscriptstyle
{}_2F_1(a,b;c;z)}&=&\scriptscriptstyle{\frac{\Gamma(b-a)\Gamma(c)(-z)^{-a}}{\Gamma(b)\Gamma(c-a)}
\sum_{n=0}^{\infty}\frac{(a)_l(a-b+1)_lz^{-l}}{l!(a-b+1)_l}}\nonumber\\
&&\scriptscriptstyle{+\frac{\Gamma(a-b)\Gamma(c)(-z)^{-b}}{\Gamma(a)\Gamma(c-b)}
\sum_{n=0}^{\infty}\frac{(b)_l(b-c+1)_lz^{-l}}{l!(b-a+1)_l}}
\eear
and $a-b$ must not be an integer. Here, for $0<n<1$ the tachyon field becomes real value
by introducing a complex constant $l$.
The warped factor $f(x)$ is~\cite{Barbosa-Cendejas:2017vgm}
\bear\label{f0}\textstyle
f(x)=-n\ln\bigg[\frac{\sigma^2\cosh(H(\frac{x}{n}+\lambda)}{H^2}\bigg].
\eear
For $n=1/2$, the potential $V(x)$ (\ref{V0}) is obtained as
\bear\textstyle
V(x)=\frac{3\sqrt{6}\,\sigma^2}{k_5^2}{\rm sech}\big[H(2x+\lambda)\big],
\eear
the tachyon field $T(x)$ (\ref{T0}) leads to
\bear{\textstyle
T(x)=\pm\frac{1}{\sqrt{2}\,\sigma}F\bigg[iH\big(x+\frac{\lambda}{2}\big),2\bigg]},
\eear
where for $-\pi/2<\phi<\pi/2$, the elliptic integral of the first kind $F(\phi,\kappa)$ is defined by
\bear\textstyle
F(\phi,\kappa)=\int_{0}^{\phi}\frac{1}{\sqrt{1-\kappa^2 \sin^2(\theta)}}d\theta.
\eear
For $n=1/2$, the warped factor $f(x)$ (\ref{f0}) is given as
\bear\textstyle
f(x)=-\frac{1}{2}\ln\bigg[\frac{\sigma^2\cosh[H(2x+\lambda)]}{H^2}\bigg].
\eear
Then, the energy density $\rho$, and the five-dimensional curvature scalar ${\cal R}$ are
\bear\textstyle
\rho(x)=\frac{6\sqrt{6}H^2}{k_5^2(\cosh[2H(2x+\lambda)]+1)},
\eear
\bear\textstyle
{\cal R}=28\sigma^2{\rm sech}[H(2 x+\lambda)].
\eear

When $\Lambda_5=0$ and $n=1/2$, shapes of potential $V(x)$ as the function of the position $x$
for various values of $H$ are depicted in Fig. 7.
The smaller $H$ becomes , the more convex function in potential $V(x)$.
Then, profiles of tachyon filed  $T(x)$ for various $H$ and
profiles of energy density are depicted in Fig. 8 and Fig. 9.
The five-dimensional curvature scalar ${\cal R}$ is depicted in Fig. 10.
The configurational entropy of tachyon filed $S_{\rm c, Flat}$ is numerically calculated
by Eqs. (\ref{den}), (\ref{mf}), and (\ref{CS}), and
is depicted in Fig. 11.
Especially, $H$ grows up, the configurational entropy
reaches the minimum value ($S_{\rm c,Flat}=0.013826$) at a critical point ($H_c=1.26$).
Thus, as discussed in the previous AdS case, it is expected that the
predominant tachyonic states happens at the minimum configurational entropy.

\begin{figure}[!htbp]
\begin{center}
{\includegraphics[width=8cm]{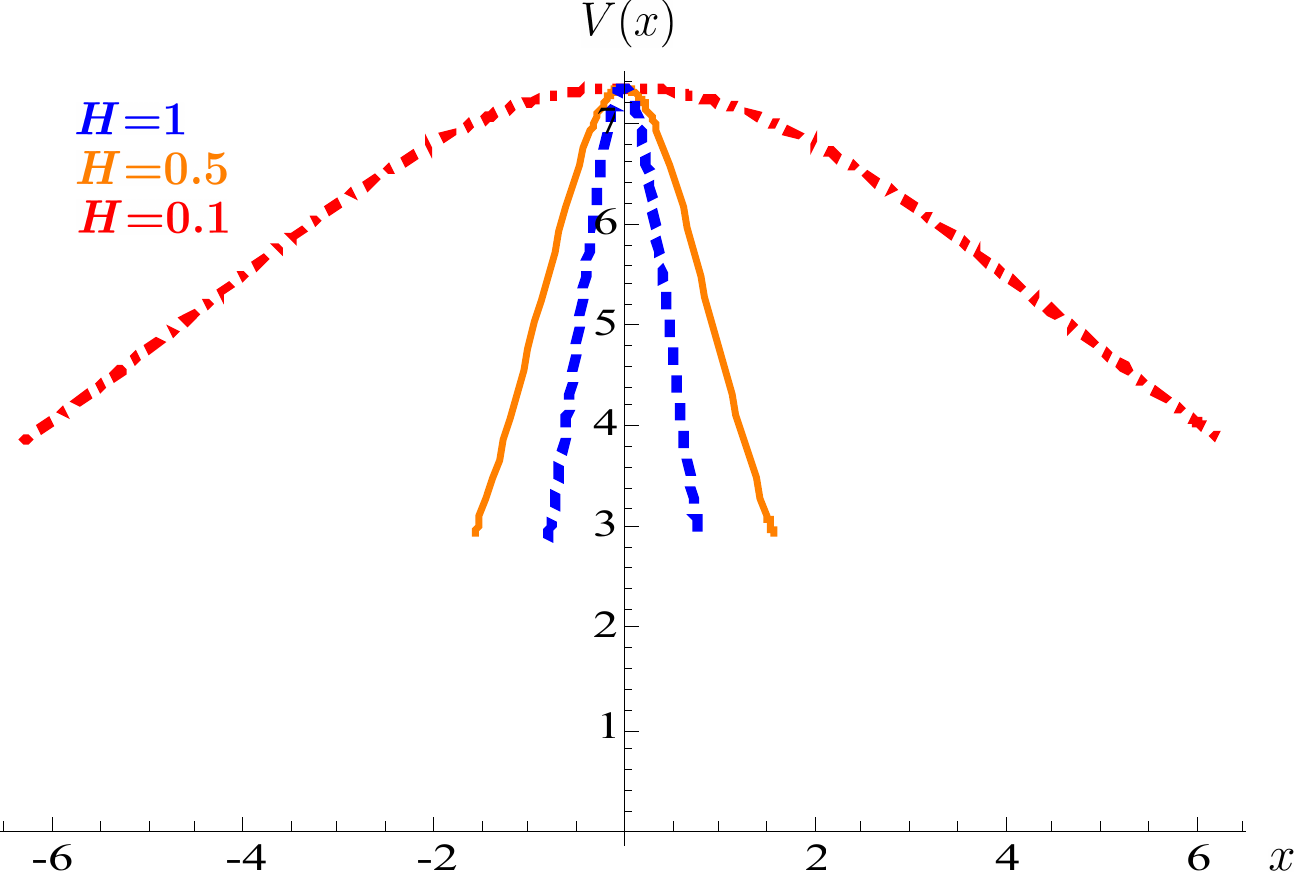}}
\end{center}
\vspace{-0.6cm}
\caption{{\footnotesize Plot of potential $V(x)$ as
the function of the position $x$
for $k_5=1$, $n=1/2$, $\sigma=1$ and $\lambda=0$
(red dotted-dashed curve for $H =0.1$, orange solid curve for $H=0.5$,
blue dashed curve for $H=1$,
respectively).}}
\label{figV}
\end{figure}

\begin{figure}[!htbp]
\begin{center}
{\includegraphics[width=8cm]{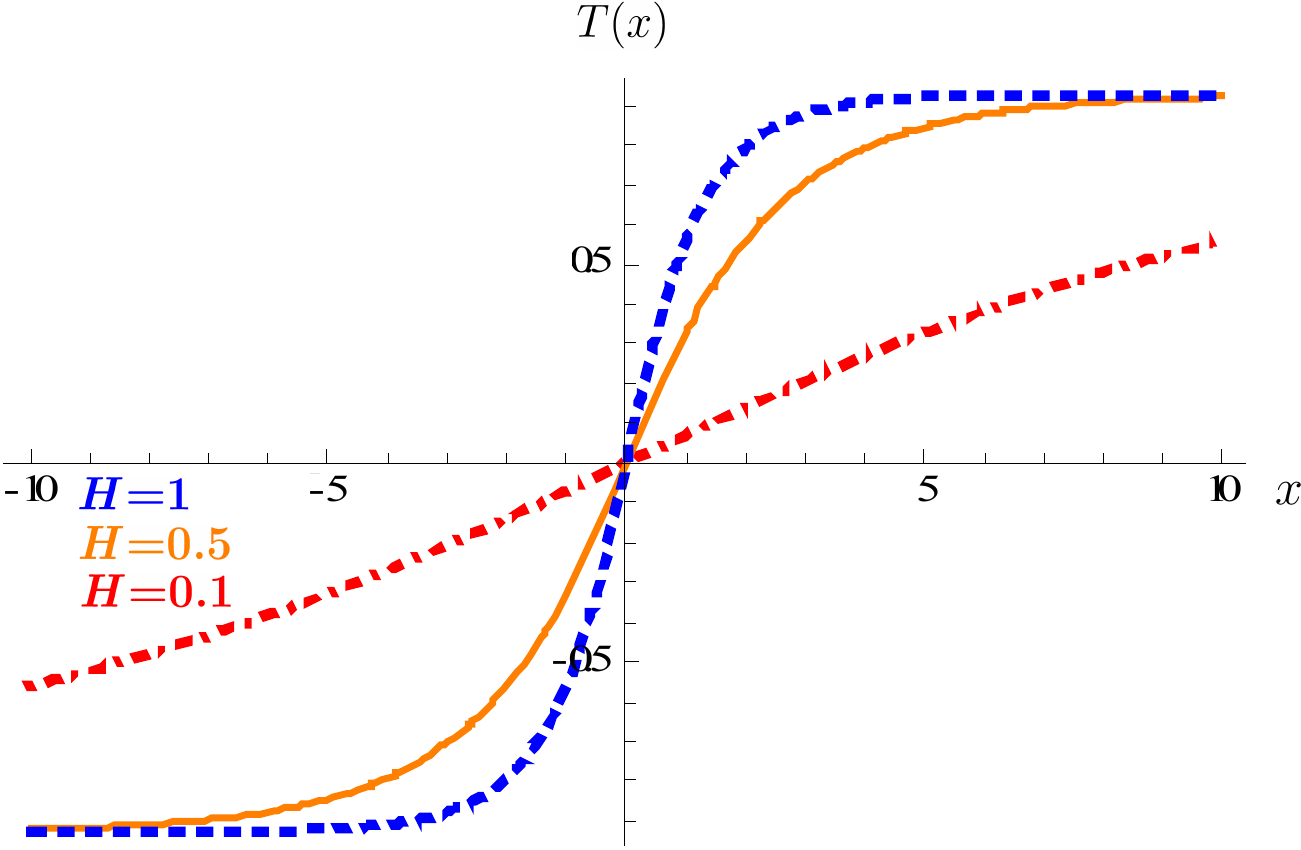}}
\end{center}
\vspace{-0.6cm}
\caption{{\footnotesize Plot of the tachyon field $T(x)$ as
the function of the position $x$
for $n=1/2$, $\sigma=1$ and $\lambda=0$
(red dotted-dashed curve for $H =0.1$, orange solid curve for $H=0.5$,
blue dashed curve for $H=1$,
respectively).}}
\label{figVI}
\end{figure}

\begin{figure}[!htbp]
\begin{center}
{\includegraphics[width=8cm]{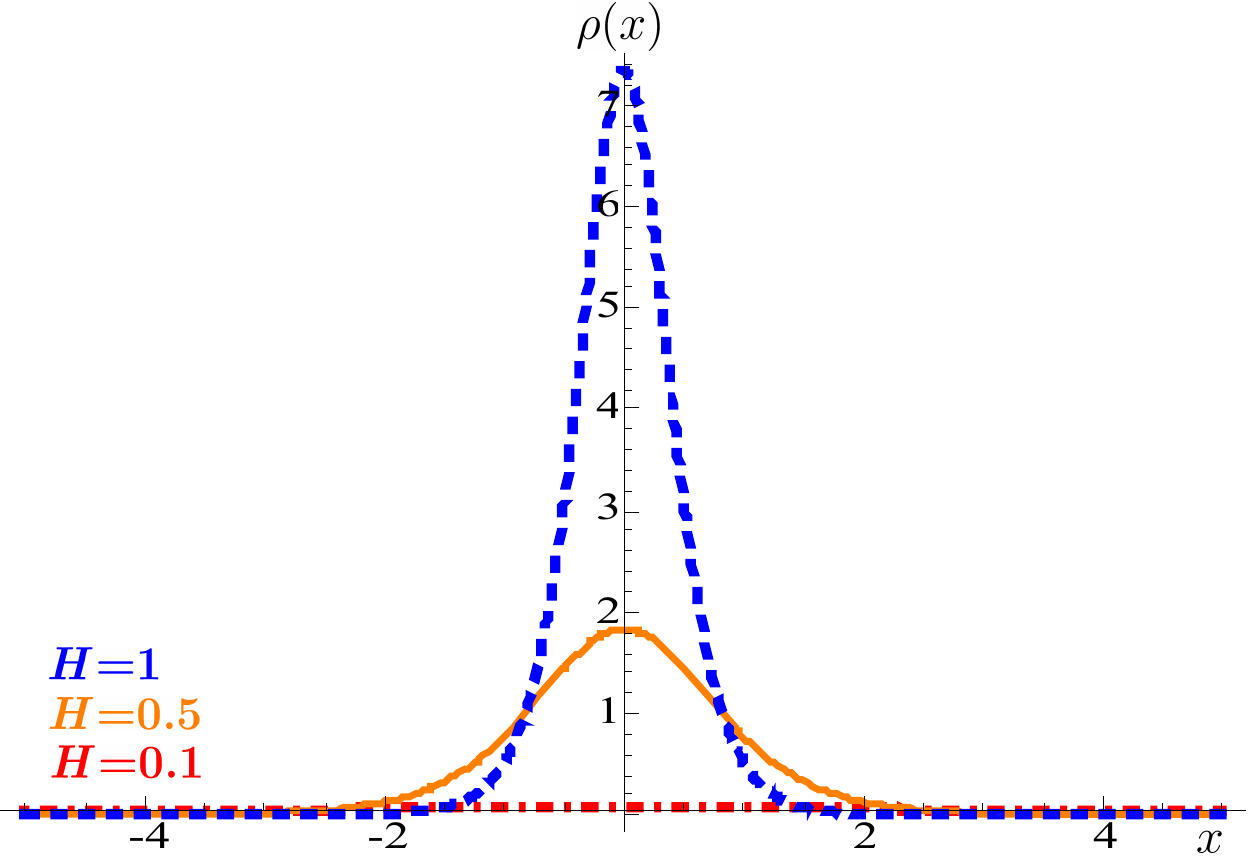}}
\end{center}
\vspace{-0.6cm}
\caption{{\footnotesize Plot of energy density $\rho(x)$ as
the function of the position $x$
for $k_5=1$, $n=1/2$, $\sigma=1$ and $\lambda=0$
(red dotted-dashed curve for $H =0.1$, orange solid curve for $H=0.5$,
blue dashed curve for $H=1$,
respectively).}}
\label{figVII}
\end{figure}

\begin{figure}[!htbp]
\begin{center}
{\includegraphics[width=8cm]{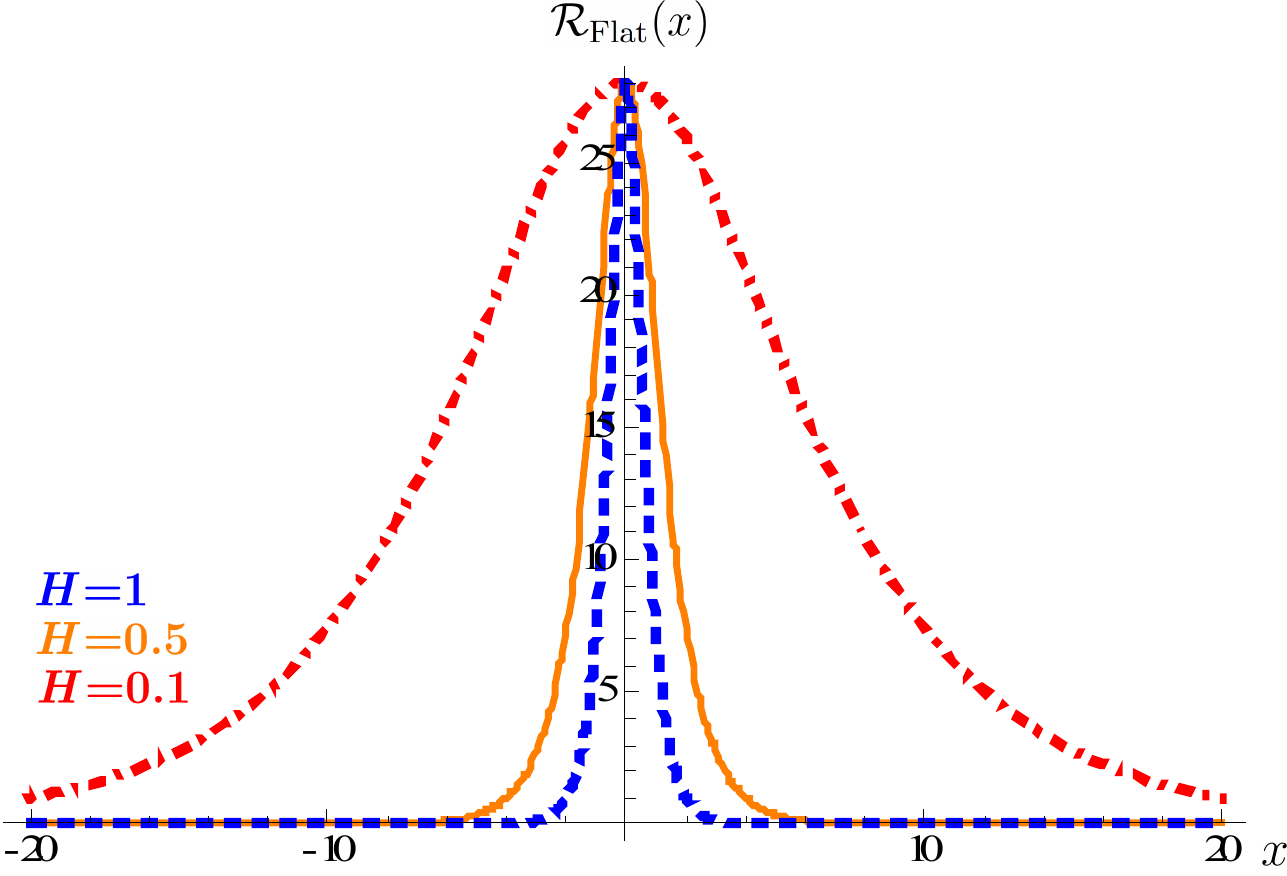}}
\end{center}
\vspace{-0.6cm}
\caption{{\footnotesize Plot of curvature sclar ${\cal R}_{\rm Flat}(x)$ as
the function of the position $x$
for $n=1/2$, $\sigma=1$ and $\lambda=0$
(red dotted-dashed curve for $H =0.1$, orange solid curve for $H=0.5$,
blue dashed curve for $H=1$,
respectively).}}
\label{fig2}
\end{figure}

\begin{figure}[!htbp]
\begin{center}
{\includegraphics[width=8cm]{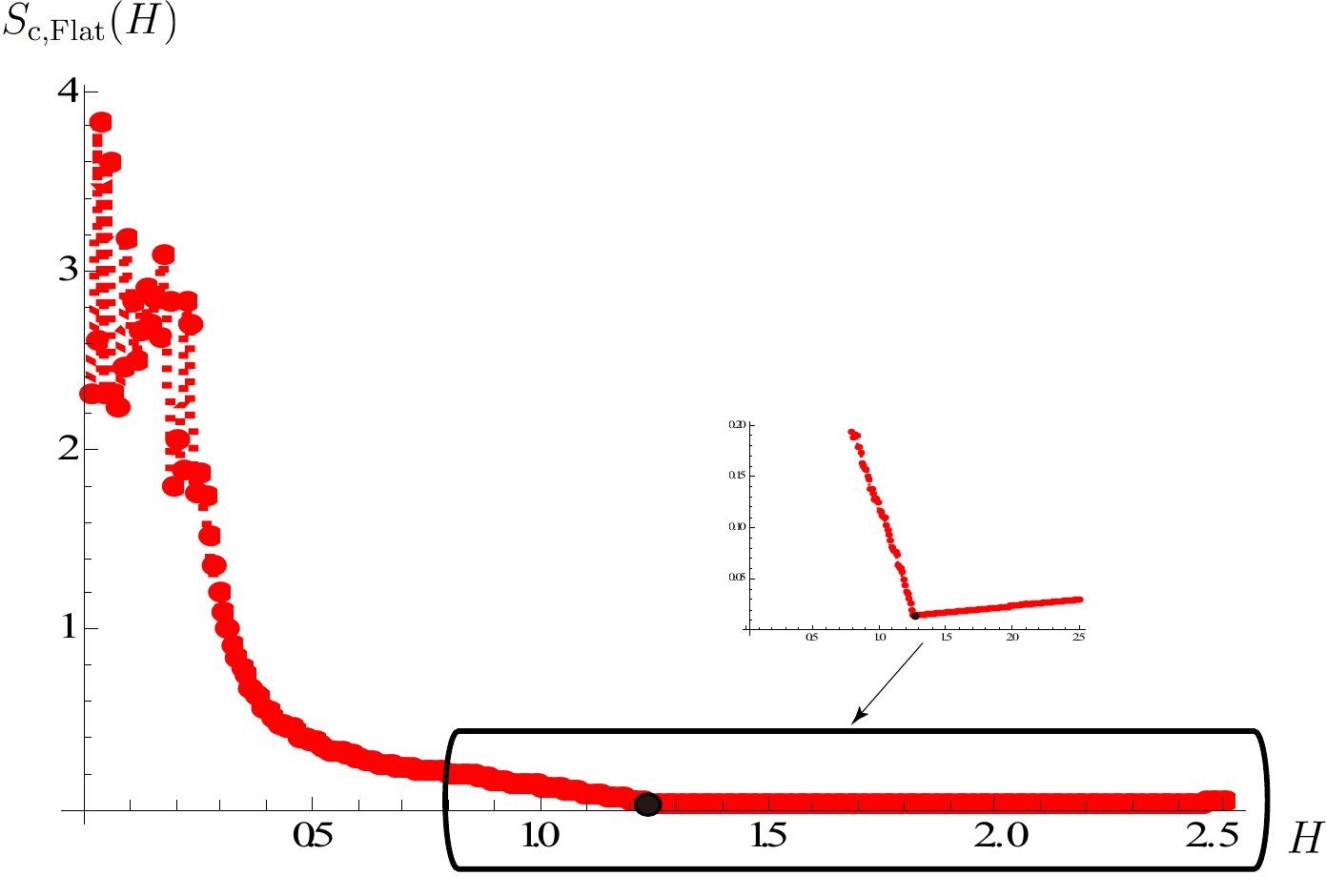}}
\end{center}
\vspace{-0.6cm}
\caption{{\footnotesize Plot of tachyon configurational entropy $S_{\rm c,Flat}$
as the function of $H$ for $k_5=1$, $n=1/2$, $\sigma=1$ and $\lambda=1$.}}
\label{figVIII}
\end{figure}

\subsection{$\Lambda_5>0$ case}
In the case of the five-dimensional gravity action with the
positive cosmological constant $\Lambda_5 >0$,
the potential $V(x)$, the tachyon field $T$, the warped factor $f(x)$, the energy density $\rho(x)$,
and the five-dimensional curvature scalar ${\cal R}$
are obtained as
\bear\textstyle{
V(x)=-\Lambda_5 \frac{\sqrt{3\sec[H(2x+\lambda)]+2}\sqrt{\sec[H(2x+\lambda)]+1}}{\sqrt{2}}},
\eear
\bear
\textstyle{
T(x)=\pm\sqrt{\frac{3}{2k_5^2 \Lambda_5}} \tanh^{-1}
\left[\frac{\sin\left(\frac{H(2x+\lambda)}{2}\right)}{\sqrt{\cos(H(2x+\lambda))}}\right]},
\eear
\bear\textstyle
f(x)=\frac{1}{2}\ln[\frac{6H^2{\rm sec}[H(2x+\lambda)]}{k_5^2 \Lambda_5}],
\eear
\bear\textstyle
\rho(x)=\frac{6H^2\sqrt{\sec[H(2x+\lambda)+1]}\sqrt{3\sec[H(2x+\lambda)+2]}}
{k_5^2(\cos[H(2x+\lambda)]+1)\sqrt{2-1/(\cos[H(2x+\lambda)]+1)}},
\eear
\bear\textstyle
{\cal R}=\frac{14}{3}k_5^2 \Lambda_5 {\rm sec}[H(2x+\lambda)].
\eear

\begin{figure}[!htbp]
\begin{center}
{\includegraphics[width=8cm]{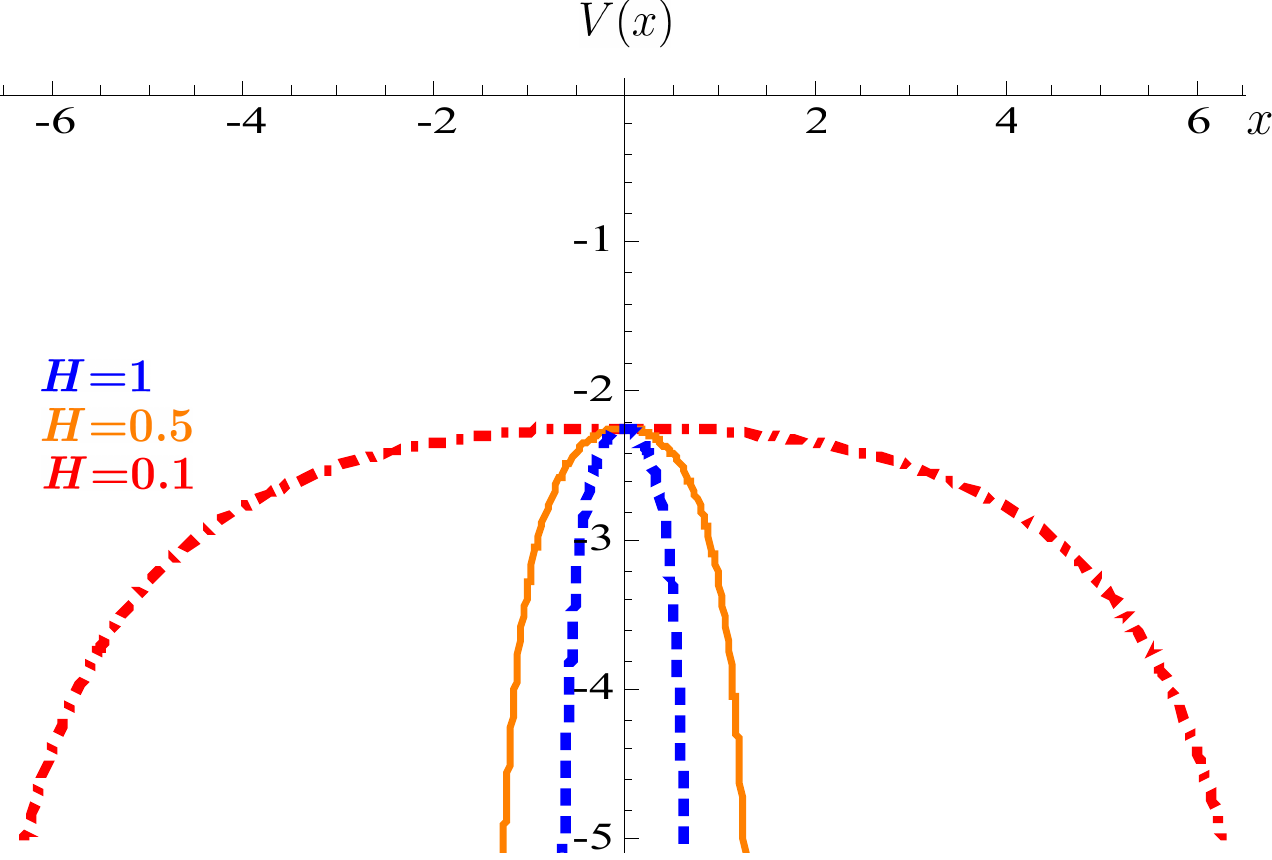}}
\end{center}
\vspace{-0.6cm}
\caption{{\footnotesize Plot of potential $V(x)$ as
the function of the position $x$
for $k_5=1$, $\Lambda_5=1$, and $\lambda=0$
(red dotted-dashed curve for $H =0.1$, orange solid curve for $H=0.5$,
blue dashed curve for $H=1$,
respectively).}}
\label{figXII}
\end{figure}

\begin{figure}[!htbp]
\begin{center}
{\includegraphics[width=8cm]{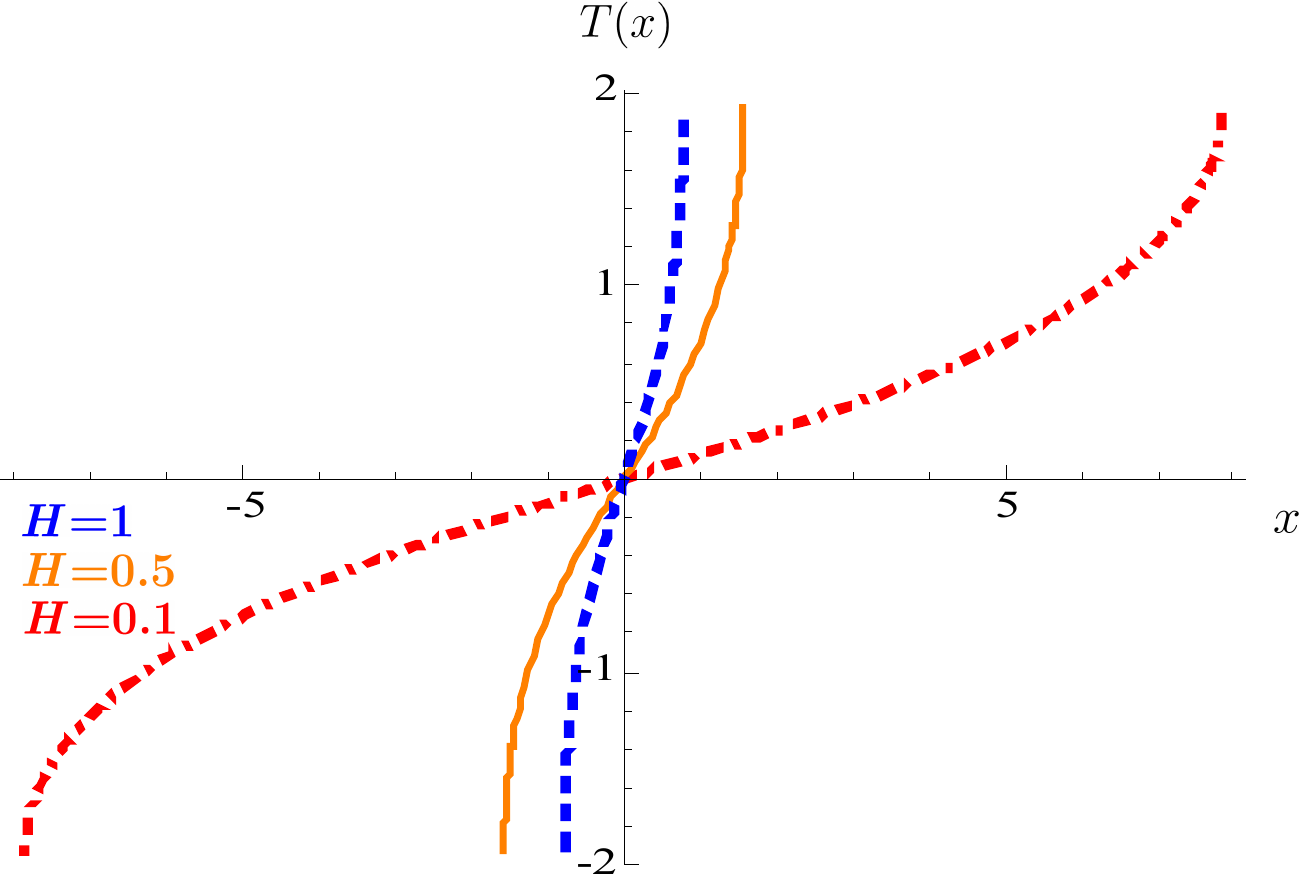}}
\end{center}
\vspace{-0.6cm}
\caption{{\footnotesize Plot of the tachyon field $T(x)$ as
the function of the position $x$
for $k_5=1$, $\Lambda_5=1$, and $\lambda=0$
(red dotted-dashed curve for $H =0.1$, orange solid curve for $H=0.5$,
blue dashed curve for $H=1$,
respectively).}}
\label{figXIII}
\end{figure}

\begin{figure}[!htbp]
\begin{center}
{\includegraphics[width=8cm]{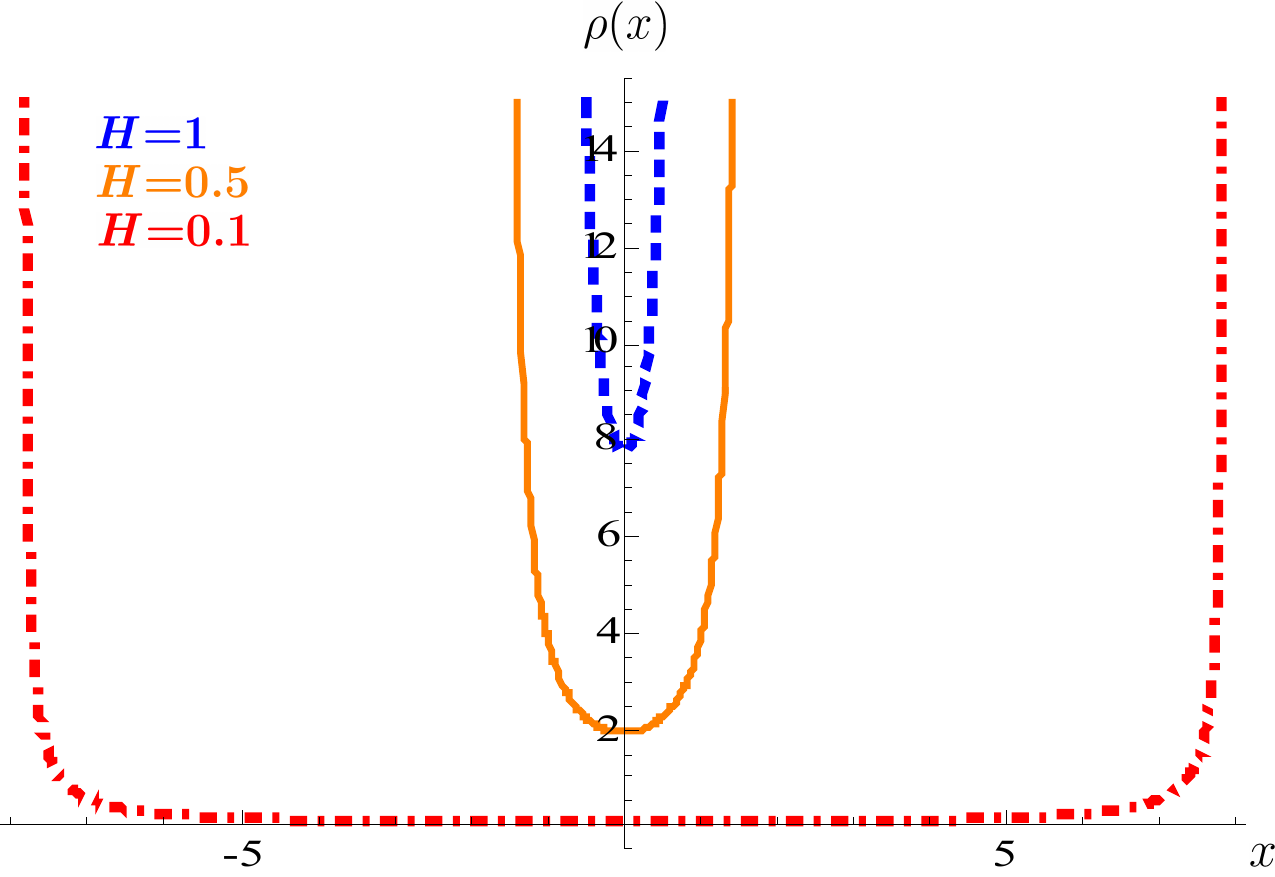}}
\end{center}
\vspace{-0.6cm}
\caption{{\footnotesize Plot of energy density $\rho(x)$ as
the function of the position $x$
for $k_5=1$, $\Lambda_5=1$, and $\lambda=0$
(red dotted-dashed curve for $H =0.1$, orange solid curve for $H=0.5$,
blue dashed curve for $H=1$,
respectively).}}
\label{figXIV}
\end{figure}

\begin{figure}[!htbp]
\begin{center}
{\includegraphics[width=8cm]{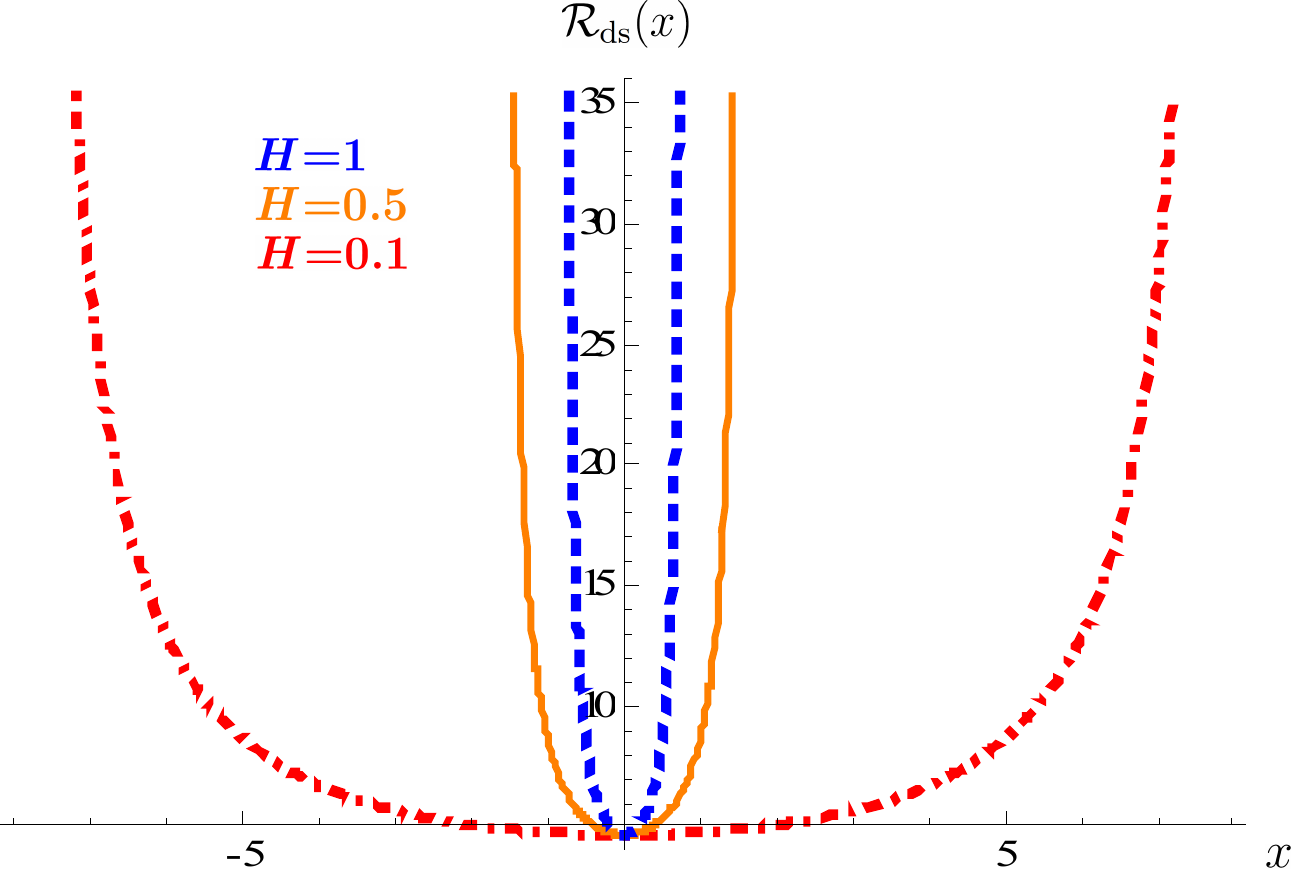}}
\end{center}
\vspace{-0.6cm}
\caption{{\footnotesize Plot of curvature scalar ${\cal R}_{\rm dS}(x)$ as
the function of the position $x$
for $k_5=1$, $\Lambda_5=1$, and $\lambda=0$
(red dotted-dashed curve for $H =0.1$, orange solid curve for $H=0.5$,
blue dashed curve for $H=1$,
respectively).}}
\label{figXV}
\end{figure}

\begin{figure}[!htbp]
\begin{center}
{\includegraphics[width=8cm]{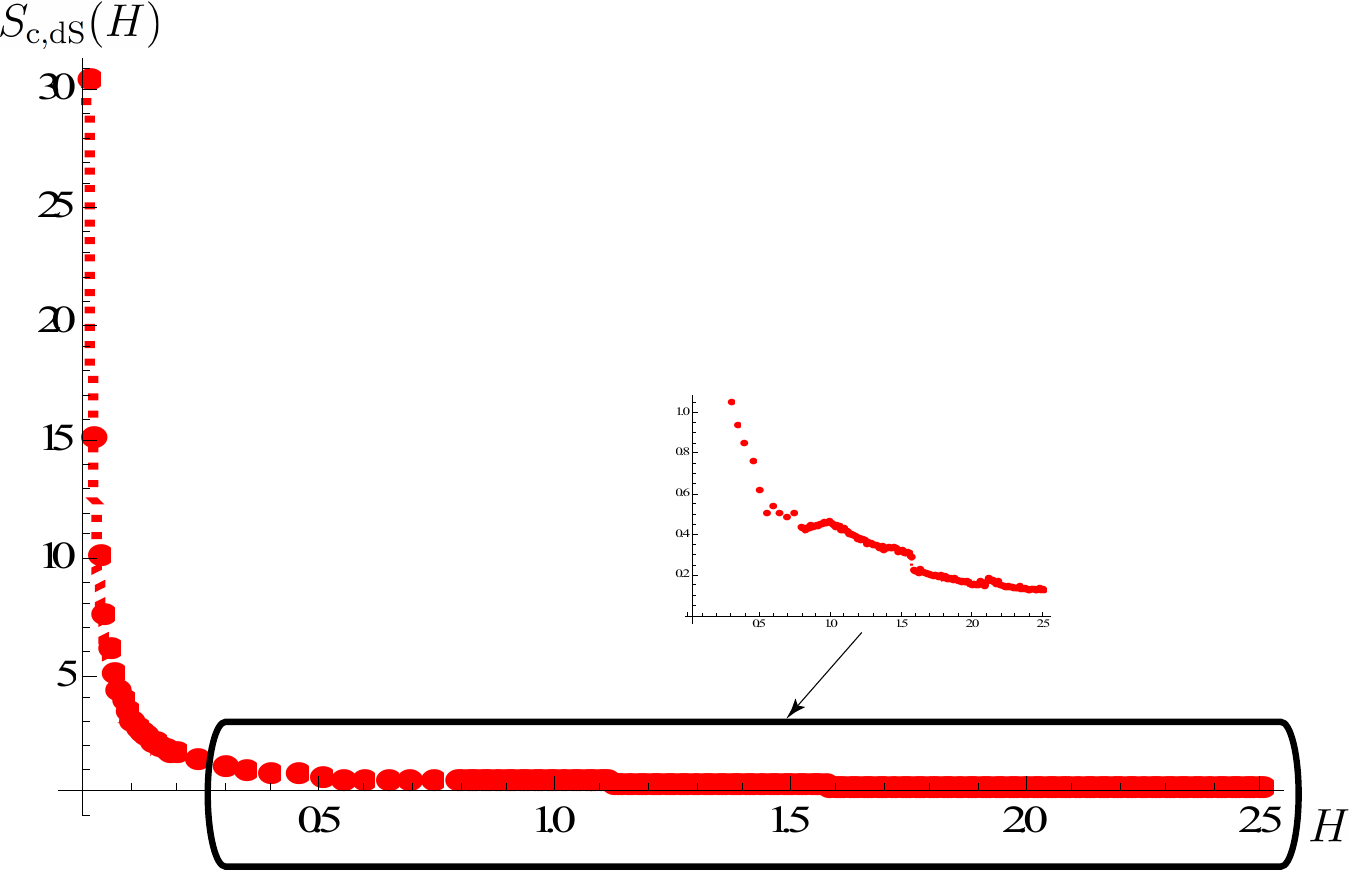}}
\end{center}
\vspace{-0.6cm}
\caption{{\footnotesize Plot of tachyon configurational entropy $S_{\rm c,dS}$
as the function of $H$ for $k_5^2=1$, $\Lambda_5=1$, and $\lambda=1$.}}
\label{figXVI}
\end{figure}

When $\Lambda_5>0$, shapes of potential $V(x)$ as the function of the position $x$
for various values of $H$ are depicted in Fig. 12.
The bigger $H$ becomes, the more convex function in potential $V(x)$.
Then, profiles of tachyon filed  $T(x)$ for various $H$ and
profiles of energy density are depicted in Fig. 13 and Fig. 14, respectively.
The five-dimensional curvature scalar ${\cal R}$ is depicted in Fig. 15.
The configurational entropy of tachyon filed $S_{\rm c, dS}$ is numerically calculated
by Eqs. (\ref{den}), (\ref{mf}), and (\ref{CS}), and is depicted in Fig. 16.
As $H$ grows up, $S_{\rm c, dS}$ almost monotonically decreases,
which contrasts with the results in Fig. \ref{figIV} and Fig. \ref{figVIII}.
Here, it seems that the local minimum configurational entropy is absent.

\section{Instability of tachyonic braneworld}
In this section, employing  superpotential operators via supersymmetric quantum mechanics,
we will investigate instability of tachyonic braneworld model under scalar fluctuations
~\cite{German:2015cna,Giovannini:2001fh, Kobayashi:2001jd,Giovannini:2002xz}.

The respective fluctuations of tachyonic braneworld model under scalar perturbations are given as the following equations
\cite{German:2012rv, German:2015cna}
\bear
\delta \Box T-\delta\left(\frac{\nabla_{\mu}\nabla_{\nu}T\nabla^{\mu}T\nabla^{\nu}T}{1+g^{\mu\nu}\nabla_{\mu}T\nabla_{\nu}T}\right)
=\delta\left(\frac{\partial_T V}{V}\right),
\eear
\bear
\delta R_{\,\mu}^{\nu}-\frac{1}{2}\delta_{\,\mu}^{\nu}\delta R=k_5^2\delta T_{\,\mu}^{\nu}.
\eear
Employing $\varphi(T)=\varphi_0(T)+\delta\varphi(T)$ and considering the scalar sector of the 5D perturbed metric
\bear
ds_5^2 = e^{2f}[(1+2\phi)dx^2+(1+2\psi)\gamma_{cd}d\alpha^{c}d\alpha^{d}],
\eear
the perturbed bulk energy-momentum tensor of the tachyonic fields is given as
\bear{\scriptscriptstyle
\delta T_{\,\mu}^{\nu}}&=&-{\scriptscriptstyle\left[\sqrt{1+e^{-2f}\varphi_0^{'2}}(\partial_{\varphi_0}V)\delta \varphi
+\frac{e^{-2f}\varphi_0^{'}V(\delta \varphi^{'}-\phi\varphi_0^{'} )}{\sqrt{1+e^{-2f}\varphi_0^{'2}}}\right]\delta_{\,\mu}^{\nu}}\nonumber\\
&&{\scriptscriptstyle +e^{-2f}\varphi_0^{'2}\left[\frac{(\partial_{\varphi_0}V)\delta \varphi}{\sqrt{1+e^{-2f}\varphi_0^{'2}}}
-\frac{e^{-2f}\varphi_0^{'}V(\delta \varphi^{'}-\phi\varphi_0^{'} )}{(1+e^{-2f}\varphi_0^{'2})^{3/2}}\right]\delta_{\, \mu}^{x}\delta_{\, x}^{\nu}}\nonumber\\
&&{\scriptscriptstyle+\frac{\varphi_0^{'}V}{\sqrt{1+e^{-2f}\varphi_0^{'2}}}[\delta_{\, \mu}^{x} \partial^{\nu}\delta \varphi
+e^{-2f}(\delta_{\, x}^{\nu}\partial_{\mu}\delta\varphi-2\delta_{\, \mu}^{x}\delta_{\, x}^{\nu}\phi\varphi_0^{'})]}
\eear
Here, the small Roman indices $(c,d =0,1,2,3.)$ denote the spacetime indices of the braneworlds.

The generalized definition of the 4D mass space-times with constant curvature is defined as
\cite{German:2015cna,Giovannini:2001fh, Kobayashi:2001jd,Giovannini:2002xz}
\bear
\gamma^{cd}\nabla_{c}\nabla_{d}-2H^2\equiv m^2,
\eear
where the Hubble parameter $H$ is connected to the spatial curvature  $K$ as follows $H^2=K$.
After some algebra, one can get the following equation
\bear
&&{\scriptscriptstyle(A-1)\psi^{''}+[f^{'}-D+B(A-2)]\psi^{'}+2[2f^{''}-2f^{'2}-f^{'}D+C(A-2)-2K]\psi}\nonumber\\
&=&{\scriptscriptstyle(\gamma^{cd}\nabla_{c}\nabla_{d}-2K)\psi\equiv m^2 \psi}
\eear
where the functions $A(x)$, $B(x)$, $C(x)$, $D(x)$, $F(x)$ are given as
$A=\frac{e^{-2f}\varphi_0^{'2}}{1+e^{-2f}\varphi_0^{'2}}$, $B=\frac{F^{'}}{F}+2f^{'}$, $C=\frac{F^{'}}{F}f^{'}+f^{''}+\frac{\varphi_0^{'}}{F}$,
$D=\frac{\varphi_0^{'}}{V}\partial_{\varphi_0}V$, $F=-\frac{3\sqrt{1+e^{-2f}\varphi_0^{'2}}}{k_5^2\varphi_0^{'}V}$,
which leads to  a Schr\"{o}dinger-like form
\bear
-(1-A)G^{''}+{\cal V}G=m^2 G,
\eear
through the following transformation $\psi=\frac{G(x,\alpha^{c})}{\sqrt{\xi(x)}}$.
Here the associated potential ${\cal V}$ is given as
\bear{\scriptscriptstyle
{\cal V}}={\scriptscriptstyle\frac{7}{2}f^{''}-4f^{'2}-\frac{1}{2}\eta_1^{'}+2\eta_2-4K}
{\scriptscriptstyle-\frac{\left[\frac{1}{2}f^{'2}+\frac{1}{2}\eta_1^2+f^{'}\eta_1-A^{'}(f^{'}+\eta_1)\right]}{2(A-1)}},\nonumber\\
\eear
with $\eta_1 =-D+B(A-2)$ and $\eta_2 =-f^{'}D+C(A-2)-4K$.
The potential ${\cal V}(x)$ is related to the superpotential ${\cal J}(x)$
\bear
{\cal V}={\cal J}^2-\sqrt{1-A}{\cal J}^{'}-\frac{1}{4}A^{''}-\frac{3}{16}\frac{A^{'2}}{1-A},
\eear
and the supersymmetric operators $\Pi^{\dag}$ and $\Pi$ are obtained as
\bear\label{pidpi1}
\Pi^{\dag}&=&\left(-\sqrt{1-A}\frac{\partial}{\partial x}-\frac{1}{4}\frac{A^{'}}{\sqrt{1-A}}+{\cal J}\right),\\
\label{pidpi2}
\Pi&=&\left(\sqrt{1-A}\frac{\partial}{\partial x}+\frac{1}{4}\frac{A^{'}}{\sqrt{1-A}}+{\cal J}\right),
\eear
which leads to
\bear\label{sle}
{\scriptstyle\Pi^{\dag}\Pi F}
&=&{\scriptstyle-(1-A)F^{''}+\left(-\frac{1}{4}A^{''}-\frac{3}{16}\frac{A^{'2}}{1-A}+{\cal J}^2-\sqrt{1-A}{\cal J}^{'}\right)F}\nonumber\\
&=& {\scriptstyle m^2F}.
\eear
Once the Schrodinger-like equation can be written in terms of the supersymmetric operators $\Pi^{\dag}$ and $\Pi$,
there are no states with usable modes, which is borrowed from supersymmetric quantum mechanics
\cite{German:2015cna,Giovannini:2001fh, Kobayashi:2001jd,Giovannini:2002xz}.
Employing the auxiliary field ${\cal X}$, the superpotential ${\cal J}$ may be defined as
\bear\label{sp}
{\cal J}=\sqrt{1-A}\left(\frac{A^{'}}{1-A}-\frac{{\cal X}^{'}}{{\cal X}}\right)
\eear
After some algebra, Eq. (\ref{sle}) reduces to
\bear\label{spe}
{\cal X}^{''}+{\cal Y}{\cal X}=0,
\eear
where the function ${\cal Y}$ is given as
\bear
{\cal Y}=\frac{1}{A-1}\left(A^{''}+\frac{2}{A-1}(\zeta_1-A^{'2})+\zeta_2\right),
\eear
with $\zeta_1=A^{'2}+(A-1)A^{'}\frac{{\cal X}^{'}}{{\cal X}}$
and $\zeta_2=-A^{''}-2A^{'}\frac{{\cal X}^{'}}{{\cal X}}-(A-1)\frac{{\cal X}^{''}}{{\cal X}}$.
Furthermore, since the tachyon field in the action (\ref{ac}) is real,
$\sqrt{1-A}$ in the supersymmetric operators (\ref{pidpi1}) and (\ref{pidpi2}) must be real,
which applies for solutions of tachyonic braneworld in the bulk dS space.
This results in the following relations
\begin{align}
1-A=\left\{
\begin{array}{cl}
&\frac{6H^2}{6H^2-k_5^2 \Lambda_5 \cosh[H(2x+\lambda)]\varphi_0^2}~~~~~(\Lambda_5<0)\\
&\\
&\frac{H^2}{H^2+\sigma^2  \cosh[H(2x+\lambda)]\varphi_0^2}~~~~~(\Lambda_5=0,~n=\frac{1}{2}).\\
&\\
&\frac{6H^2}{6H^2+k_5^2 \Lambda_5 \cos[H(2x+\lambda)]\varphi_0^2}~~~~~(\Lambda_5>0)
\end{array}
\right.
\end{align}
When $\Lambda_5<0$, since $1\leq \cosh[H(2x+\lambda)]\leq \infty$, the range of $1-A$ is given as
$0\leq 1-A \leq \frac{6H^2}{6H^2-k_5^2 \Lambda_5 \varphi_0^2}$,
which is always positive definite.
For example, (i) taking $H=\frac{1}{\sqrt{2}}$, $k_5=\sqrt{6}$, $\Lambda_5=-1$, $\lambda=0$, and $\varphi_0=1$, $1-A$ is given as
$\frac{1}{1+2\cosh(\sqrt{2}x)}$ (see red dotted-dashed curve in Fig. 17).
(ii) taking $H=1$, $k_5=\sqrt{6}$, $\Lambda=-1$, $\lambda=0$, and $\varphi_0=1$,
$1-A$ is $\frac{1}{2\cosh^2(x)}$ (see orange solid curve in Fig. 17).
(iii) taking $H=\sqrt{10}$, $k_5=\sqrt{6}$, $\Lambda=-1$, $\lambda=0$,
and $\varphi_0=1$, $1-A$ is $\frac{10}{10+\cosh(2\sqrt{10}x)}$ (blue dashed curve in Fig. 17).
The magnitude of $1-A$ is depicted in Fig. 17, and is positive in any case.
The tachyon braneworld in the bulk AdS space is stable under scalar fluctuations,
which is consistent with that in \cite{German:2012rv, German:2015cna}.

\begin{figure}[!htbp]
\begin{center}
{\includegraphics[width=8cm]{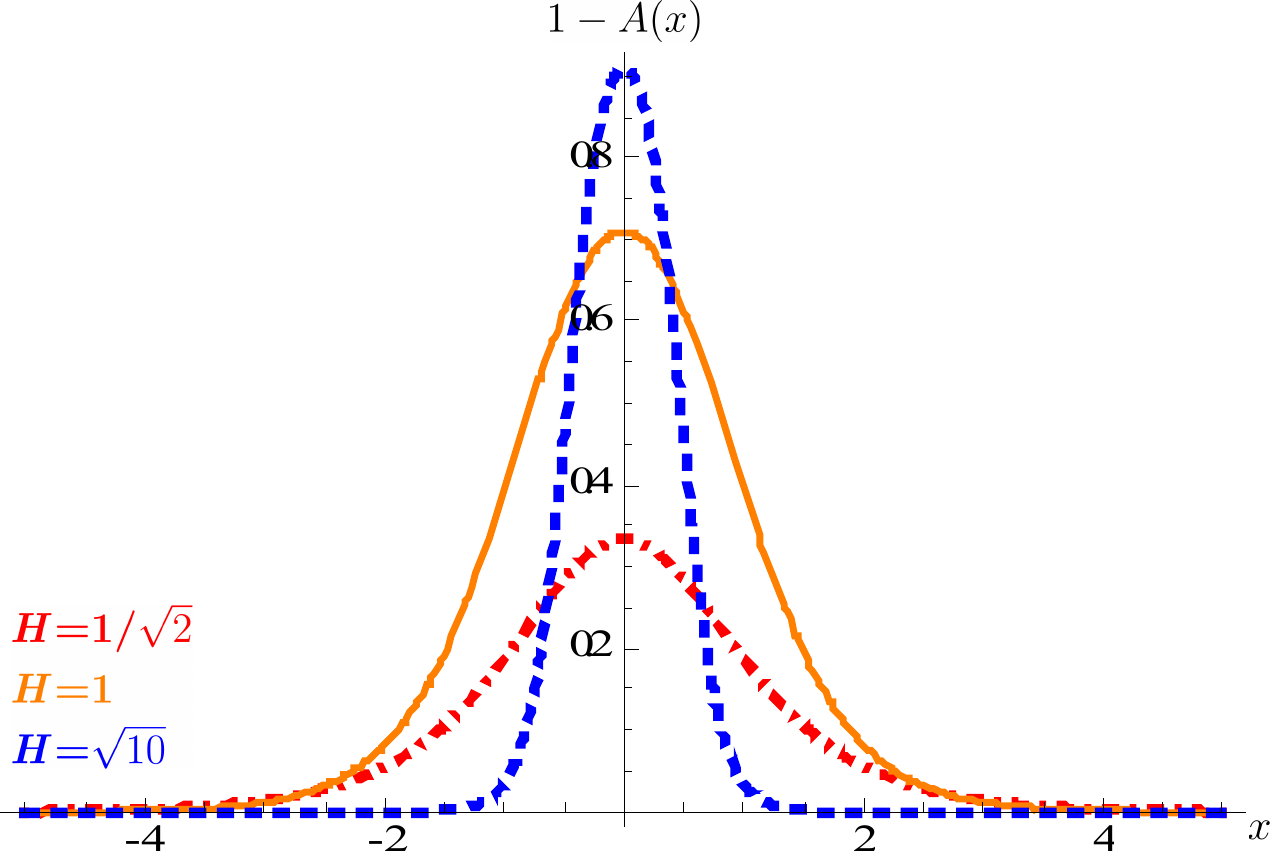}}
\end{center}
\vspace{-0.6cm}
\caption{{\footnotesize Plot of the magnitude of $1-A$
as the function of the position $x$ for $k_5=\sqrt{6}$,
$\Lambda_5=-1$, $\lambda=0$, and $\varphi_0=1$
(red dotted-dashed
curve for $H=1/\sqrt{2}$, orange solid curve for
$H=1$, blue dashed curve for $H=\sqrt{10}$, respectively).}}
\label{figXVII}
\end{figure}

When $\Lambda_5=0$, $H$ and $\sigma$ of $1-A$ are replaced by $6H$ and $-k_5^2\Lambda_5$ respectively,
which reduces to $1-A$ in the above bulk AdS case. Thus, the tachyon braneworld model without
the bulk cosmological constat is also stable under scalar fluctuations,
which is consistent with that in~\cite{Barbosa-Cendejas:2017vgm}.

\begin{figure}[!htbp]
\begin{center}
{\includegraphics[width=8cm]{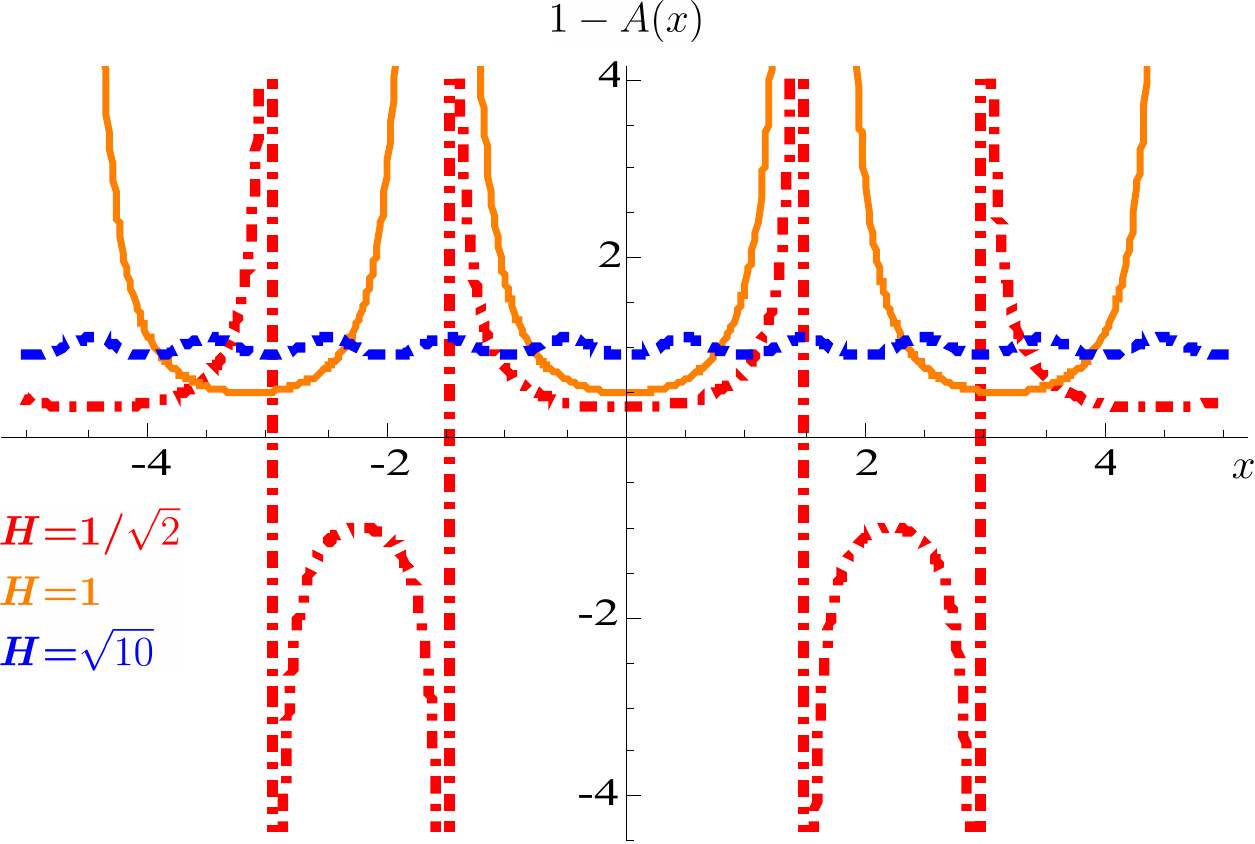}}
\end{center}
\vspace{-0.6cm}
\caption{{\footnotesize Plot of the magnitude of $1-A$
as the function of the position $x$ for $k_5=\sqrt{6}$,
$\Lambda_5=1$, $\lambda=0$, and $\varphi_0=1$
(red dotted-dashed
curve for $H=1/\sqrt{2}$, orange solid curve for
$H=1$, blue dashed curve for $H=\sqrt{10}$, respectively).}}
\label{figXVIII}
\end{figure}

Finally, when $\Lambda_5>0$, since $-1\leq\cos[H(2x+\lambda)]\leq1$,
the magnitude of $1-A$ becomes positive for $6H^2 > k_5^2\Lambda_5\varphi_0^2$.
For example, (i) taking $H=\frac{1}{\sqrt{2}}$, $k_5=\sqrt{6}$, $\Lambda_5=1$, $\lambda=0$, and $\varphi_0=1$,
in the case of $\big[(6H^2=3)< (k_5^2\Lambda_5\varphi_0^2=6)\big]$, $1-A$ is given as
$\frac{1}{1+2\cos(\sqrt{2}x)}$.
The value of the function is positive for $\frac{\sqrt{2}(3n-1)\pi}{3}<x<\frac{\sqrt{2}(3n+1)\pi}{3}$ whereas it is negative
for $\frac{\sqrt{2}(3n+1)\pi}{3}<x<\frac{\sqrt{2}(3n+2)\pi}{3}$ where $n$ is  a integer value
(see red dotted-dashed curve in Fig. 18).
(ii) taking $H=1$, $k_5=\sqrt{6}$, $\Lambda_5=1$, $\lambda=0$, and $\varphi_0=1$,
in the case of $\big[(6H^2=6)=(k_5^2\Lambda_5\varphi_0^2=6)\big]$,
$1-A$ is $\frac{1}{2\cos^2(x)}$, which is positive but is singular at some points ($x=n\pi$)
(see orange solid curve in Fig. 18).
(iii) taking $H=\sqrt{10}$, $k_5=\sqrt{6}$, $\Lambda_5=1$, $\lambda=0$, and $\varphi_0=1$,
in the case of $\big[(6H^2=60)>(k_5^2\Lambda_5\varphi_0^2=6)\big]$,
$1-A$ is $\frac{10}{10+\cos(2\sqrt{10}x)}$ and is always positive (blue dashed curve in Fig. 18).
It implies that the tachyon braneworld in the bulk dS space is stable under scalar fluctuations
when $6H^2 > k_5^2\Lambda_5\varphi_0^2$.

\newpage
\section{Conclusion}
We considered the tachyonic system coupled to gravity with the bulk cosmological constant
and investigated its configurational entropy. For $\Lambda_5<0$ or $\Lambda_5=0$,
we found that the configurational entropy becomes a global minimum
as the magnitude of scale factor reaches the critical value.
It seems that an accelerated rate of the universe and cosmological inflation rate for radiation/matter domination
can be determined by such critical value.
For $\Lambda_5>0$, we also found that an exact solution of the tachyonic braneworld, and
the configurational entropy monotonically decreases when increasing the magnitude of scale factor.

Employing the replacement of the supersymmetric operators by the specific solution of the field equations,
we found that tachyon braneworld in the bulk negative/zero cosmological constant is stable
under scalar fluctuations, which is consistent with that in~\cite{German:2012rv,Barbosa-Cendejas:2017vgm,German:2015cna}.
We applied similar analysis to the instability
of tachyonic braneworld model in the bulk positive cosmological constant.
It was shown that such model is stable under scalar fluctuations when $6H^2 > k_5^2\Lambda_5\varphi_0^2$.

\section*{Acknowledgements}
This work was supported by Basic Science Research Program through
the National Research Foundation of Korea (NRF) funded by the
Ministry of Education, Science and Technology (NRF-2018R1D1A1B07049451).

\end{document}